\documentclass[preprintnumbers,prd,letterpaper,nofootinbib]{revtex4}

\usepackage[pdftex]{graphicx}
\usepackage{latexsym,amsmath,amssymb,lmodern,float,url}
\usepackage{natbib}
\usepackage[pdftex,bookmarks,linktocpage,pdfpagelabels,plainpages=false,hyperfigures,linkcolor=blue,citecolor=blue]{hyperref} %for hypertext links
\usepackage{verbatim}
\usepackage{xspace}
\usepackage{hyperref}
\usepackage{slashed}
\usepackage{color}
\hypersetup{colorlinks=true}

\newcommand\be{\begin{equation}}
\newcommand\ee{\end{equation}}
\newcommand\bea{\begin{eqnarray}}
\newcommand\eea{\end{eqnarray}}
\newcommand{\appropto}{\mathrel{\vcenter{
\offinterlineskip\halign{\hfil$##$\cr
\propto\cr\noalign{\kern2pt}\sim\cr\noalign{\kern-2pt}}}}}

\begin{document}

MI-TH-1636, WSU-HEP-1609

\title{Planck Constraint on Relic Primordial Black Holes}

\author{Steven Clark$^{1}$}
\author{Bhaskar Dutta$^{1}$}
\author{Yu Gao$^{1,2}$}
\author{Louis E. Strigari$^{1}$}
\author{Scott Watson$^{3}$}

\affiliation{
$^{1}$~Department of Physics and Astronomy, Mitchell Institute for Fundamental Physics and Astronomy, Texas A\&M University, College Station, TX 77843-4242, USA\\
$^{2}$~Department of Physics and Astronomy, Wayne State University, Detroit, 48201, USA\\
$^{3}$~Department of Physics, Syracuse University, Syracuse, NY 13244, USA\\
}

\begin{abstract}
We investigate constraints on the abundance of primordial black holes (PBHs) in the mass range $10^{15}-10^{17}$ g using data from the Cosmic Microwave Background (CMB) and MeV extragalactic gamma-ray background (EGB). Hawking radiation from PBHs with lifetime greater than the age of the universe leaves an imprint on the CMB through modification of the ionization history and the damping of CMB anisotropies. Using a model for redshift dependent energy injection efficiencies, we show that a combination of temperature and polarization data from Planck provides the strongest constraint on the abundance of PBHs for masses $\sim 10^{15}-10^{16}$ g, while the EGB dominates for masses $\gtrsim 10^{16}$ g. Both the CMB and EGB now rule out PBHs as the dominant component of dark matter for masses $\sim 10^{16}-10^{17}$ g. Planned MeV gamma-ray observatories are ideal for further improving constraints on PBHs in this mass range. 
\end{abstract}

\maketitle

\section{Introduction}
\label{introduction}

Shortly after the Big Bang, large density fluctuations in the early Universe may have resulted in formation of primordial black holes (PBHs)~\cite{bib:pbh}. There is a wide range of allowed masses for PBHs. Depending on the epoch and conditions during formation, PBH masses can be anywhere from approximately a gram to a million Solar masses. While low mass PBHs would have already evaporated through Hawking radiation~\cite{Hawking:1974rv}, large ones with masses $M_{\rm{BH}}>5\times10^{14}$ g would still be present today. It is also possible to have prolonged PBH formation during a non-radiation-dominated phase of the Universe where PBHs can form with a continuum mass distribution, rather than mostly at one particular mass scale as in the conventional radiation dominated case~\cite{Georg:2016yxa,Harada:2016mhb}. Stable PBHs can be cosmologically-significant, and may serve as an ideal dark matter candidate~\cite{Carr:2009jm}. 

Depending on the PBH abundance, Hawking radiation from PBHs with lifetime longer than the age of the universe may be observable. Extragalactic gamma rays strongly constrain PBHs in the mass range $\sim 10^{15} - 10^{17}$ g~\cite{Carr:2009jm}. PBHs in the mass range $\sim 10^{17} - 10^{20}$ g are bounded by femtolensing of gamma-ray bursts~\cite{Nemiroff:2001bp}, and PBHs in the mass range $\sim 10^{-10} - 10$ M$_\odot$ are constrained by gravitational microlensing~\cite{Griest:2013esa}. In addition, PBHs with mass $\gtrsim 10$ M$_\odot$ may be constrained by the accretion of matter in the early universe~\cite{Ricotti:2007au,Chen:2016pud,Ali-Haimoud:2016mbv}. Black holes at approximately this mass are constrained by X-rays observations~\cite{Gaggero:2016dpq}. For recent comprehensive reviews on astrophysical constraints on PBHs see Refs.~\cite{Carr:2016drx,Green:2016xgy}.

In this paper we present a new bound on PBHs in the mass range $\sim 10^{15} - 10^{17}$ g using the most recent Planck Cosmic Microwave Background (CMB) data~\cite{Aghanim:2015xee}. The CMB is sensitive to additional sources of energy injection during the recombination epoch, which leads to damping of the anisotropies. For PBHs, this energy injection is due to Hawking radiation. As we show, the Planck data now place a stronger bound on PBHs over a larger fraction of this mass range than previous most stringent bounds derived from the extragalactic gamma-ray background (EGB). Previous authors have used Planck data to bound PBHs in the mass regime that we study~\cite{Poulin:2016anj}; as discussed below we precisely identify the mass regime over which the CMB and EGB bounds are dominant. In addition we note that our analysis is distinct from previous studies that used early-time distortions of the CMB to bound PBHs in the mass range $10^{11} - 10^{12}$ g~\cite{Tashiro:2008sf}. 

The theoretical formalism that we utilize to constrain PBHs is similar to that used to constrain dark matter annihilation or decay~\cite{Zhang:2007zzh,Madhavacheril:2013cna,Slatyer:2015jla,Slatyer:2015kla,Liu:2016cnk}.  From the perspective of the CMB, PBH evaporation is most similar to dark matter decay in that the energy injection only depends on the PBH mass and abundance, and is at a steady rate to the present time. This energy injection can have a significant impact on ionization at low redshift. However unlike the energy from dark matter, which can be injected in the form of heavy Standard Model particles, PBHs with mass $> 10^{15}$g mostly radiate in electrons and photons, or other (near) massless species, but generally not into much heavier particle species. Because PBHs with mass  $>10^{18}$g are too cold to emit electrons, their injection into the CMB is unobservable, and the bounds are weak above this mass. 

This paper is organized as follows. In Section~\ref{blackholes} we briefly discuss the injection of radiation from PBHs. In Section~\ref{IGM_interactions} we discuss the impact of this injected energy on the IGM. In Section~\ref{recomb_hist}, we show the modifications on the ionization history, and in Section~\ref{CMB_alter} we discuss CMB distortions. In Section~\ref{black_hole_constr} we show the resulting constraints on PBHs, and Section~\ref{conclusions} presents the discussion and conclusions. 

\section{Blackhole Radiation Properties} \label{blackholes}
Depending on their mass, PBHs radiate a spectrum of particles, which decay via cascades into photons, electrons, protons, and neutrinos. These particles then deposit energy into the IGM. The injection of energy is described by the equation
\begin{equation}
\begin{aligned}
\dfrac{dE}{dVdt} & = \dot{M}_{\rm{BH}} c^{2} n_{\rm{BH}}\\
& = \frac{\dot{M}_{\rm{BH}}}{M_{\rm{BH}}} \rho_{c} c^{2} \Omega_{\rm{BH}}(z)\\
& = \frac{\dot{M}_{\rm{BH}}}{M_{\rm{BH}}} \rho_{c} c^{2} \Omega_{\rm{BH}} (1+z)^{3},
\end{aligned}
\label{equ:bh-energy}
\end{equation}
where $n_{\rm BH}$ is the PBH number density, $\rho_{c}$ is the critical density of the Universe today, $M_{\rm{BH}}$ is the PBH mass, $\Omega_{\rm{BH}}$ is the PBH density observed today relative to the critical density. 

Note that the above equations assume that PBHs are comprised of a single mass and the mass does not changes as it radiates. This is satisfied as long as the lifetime is large compared to the age of the universe and is satisfied by the masses considered in this work. Apart from some cosmetic differences, Equation (\ref{equ:bh-energy}) is identical to that for decaying dark matter~\cite{Slatyer:2016qyl}.

In order to evaluate Equation (\ref{equ:bh-energy}), an expression for $\dot{M}_{\rm{BH}}$ is required. To obtain this, we start from the fact that Hawking radiation equates the radiation from a black hole to the blackbody radiation of an object with temperature
\begin{equation}
T_{\rm{BH}}= \frac{1}{8 \pi G M} = 1.06 \rm{TeV} \times \frac{10^{10} \rm{g}}{M_{\rm{BH}}}
\label{equ:Hawking-Temp}
\end{equation}
and with an emission spectrum
\begin{equation}
\dfrac{dN}{dEdt} \propto \frac{\Gamma_s}{e^{E/T_{\rm{BH}}}-(-1)^{2s}}
\label{equ:Hawking-Dist}
\end{equation}
where $s$ is the spin of the radiated particle and $\Gamma_s$ is the absorption coefficient for the particle. For low $T_{\rm{BH}}$ the absorption coefficient can deviate greatly from the geometric optic limit~\cite{MacGibbon:1991tj}, 
\begin{equation}
\Gamma_s (M,E) = \frac{27G^2M^2E^2}{\hbar^2c^6}.
\label{equ:Gamma_s_high_limit}
\end{equation}
For Standard Model particles, the average radiated energy in the massless limit is~\cite{Carr:2009jm}
\begin{equation}
\begin{tabular}{ccc}
$E_\gamma = 5.71 T_{\rm{BH}},$ & $E_\nu = 4.22 T_{\rm{BH}},$ & $E_e = 4.18 T_{\rm{BH}}$. 
\end{tabular}
\label{equ:average-energy}
\end{equation}
For PBHs with masses in the range $10^{15} - 10^{17}$ grams, the fraction of emitted particles of different spins is~\cite{MacGibbon:1991tj}
\begin{equation}
\renewcommand{\arraystretch}{1.3}
\begin{tabular}{ccc}
$f_0 = 0.267,$ & $f^\gamma_1 = 0.06,$ & $f_{3/2} = 0.02,$\\
$f^g_2 = 0.007,$ & $f^\nu_1/2 = 0.147,$ & $f^{e\pm}_{1/2} = 0.142.$
\end{tabular}
\label{equ:bh_radiation_frac}
\end{equation}
It should be noted that these fractions are not normalized to unity, but rather to a $10^{17}$ g PBH.

With this information the energy injection in Equation (\ref{equ:bh-energy}) can be evaluated after using~\cite{MacGibbon:1991tj,Carr:2009jm}
\begin{equation}
\dot{M}_{\rm{BH}} = -5.34\times10^{25}\textrm{g}^3 \left( \sum_i{f_i} \right) M_{\rm{BH}}^{-2} \, \rm{s}^{-1}.
\label{equ:Mdot}
\end{equation}
Since only the electrons and photons interact electromagnetically for the PBH masses considered, these are the only fractions that need to be considered for calculating the energy output.

\section{Medium Interactions} \label{IGM_interactions}
The energy deposited into the IGM by PBHs is absorbed through multiple channels. Following previous studies, here we consider three channels for the IGM interaction: Hydrogen ionization, Lyman-Alpha excitations, and heating the IGM~\cite{Slatyer:2012yq,Belotsky:2014twa,Liu:2016cnk}. These effects alter the cosmological recombination equations as 
\begin{eqnarray}
\dfrac{dx_e}{dz} &=& \left( \dfrac{dx_e}{dz} \right)_{\rm{orig}} - \quad \frac{1}{(1+z)H(z)} (I_{Xi} (z) + I_{X\alpha} (z))
\label{equ:dxe_dz} \\
\dfrac{dT_{\rm{IGM}}}{dz} &=& \left( \dfrac{dT_{\rm{IGM}}}{dz} \right)_{\rm{orig}} - \quad \frac{2}{3 k_B (1+z)H(z)}\frac{K_h}{1+f_{\rm{He}}+x_e},
\label{equ:dT_dz}
\end{eqnarray}
where $f_{\rm{He}}$ is the helium fraction, $x_e$ is the ionization fraction, $k_B$ is the Boltzmann constant, and $H(z)$ is the Hubble parameter. The standard equations without additional energy injection from PBHs are denoted by the subscript ``orig" and are derived in e.g. Ref.~\cite{AliHaimoud:2010dx}. The quantities $I_{Xi}(z)$, $I_{X\alpha}(z)$, and $K_h$ are factors corresponding to the additional energy injection affecting ionization from the ground state, ionization from excited states, and heating the IGM. Each of these injections are dependent on the injection energy through
\begin{eqnarray}
I_{Xi} (z) &=& f_i(E,z) \frac{dE/dV dt}{n_H (z) E_i} 
\label{equ:I_Xi} \\
I_{X\alpha} (z) &=& f_\alpha(E,z) (1-C) \frac{dE/dV dt}{n_H (z) E_\alpha}
\label{equ:I_Xalpha} \\
K_{h} (z) &=& f_h(E,z) \frac{dE/dV dt}{n_H (z)}. 
\label{equ:K_h}
\end{eqnarray}
Here $n_H$ is the hydrogen number density, and $E_i$ and $E_\alpha$ are the energies of the ground and the excited hydrogen atom electron levels respectfully. The quantity $C$ is related to the probability for an excited hydrogen atom to emit a photon prior to being ionized~\cite{Madhavacheril:2013cna}. The quantities $f_i(E,z)$, $f_\alpha(E,z)$, $f_h(E,z)$ are efficiencies for energy interactions through each channel. Commonly referred to as effective efficiencies, they are redshift, energy, and species dependent quantities that equate the total energy injection to the actual amount absorbed through a pathway~\cite{Liu:2016cnk}. Previously, these efficiencies have been approximated by simple $x_e$ dependent equations, with the energy injection taken to be instantaneous, through a technique known as the ``SSCK'' method, described in further detail in Ref.~\cite{Madhavacheril:2013cna,Slatyer:2015jla}.

To calculate the effective efficiencies, we follow the approach adopted in Ref.~\cite{Slatyer:2015kla,Liu:2016cnk}. These efficiencies have been tabulated for electron and photon particle injection into the IGM at various redshifts and particle energies, and divided into five different channels: Hydrogen ionization, Helium ionization, Lyman-alpha excitations, heating, and continuum photons (energy lost as photons with $E < 10.2 \, \rm{eV}$) \cite{Slatyer:2015kla}. The Hydrogen, Lyman-alpha, and heating efficiencies were used to calculate the various efficiencies in Equations (\ref{equ:I_Xi}), (\ref{equ:I_Xalpha}), and (\ref{equ:K_h}). We do not consider Helium ionization because it is subdominant. 

We additionally note that we do not consider energy deposited into continuum photons. This is because continuum photon energy affects the CMB mostly through spectral distortions rather than through anisotropies, and the anisotropies are the focus of this paper. The impact on spectral distortions is also less significant than on anisotropies~\cite{Slatyer:2015kla}; a basic demonstration of this is given in the Appendix.

Refs.~\cite{Slatyer:2015kla,Liu:2016cnk} tabulate efficiencies specifically for decaying and annihilating dark matter into electron and photon channels. These efficiencies are specifically valid for a small ionization fraction, in which case the efficiency scales linearly with ionization fraction. For dark matter decay, the energy injection is linearly proportional to the density, and for annihilation it is proportional to the density-squared. From Equation~\ref{equ:bh-energy}, the radiation from PBHs is linearly proportional to density, therefore this is most analogous to dark matter decay. For this reason, we utilize the decay efficiencies to produce black hole efficiencies. To obtain the PBH efficiency, we assume that PBH injected particles have the same efficiency as those with the average energy for its species. Combining the different species efficiencies through a weighted average based upon their emission fractions gives 
\begin{equation}
f_{\rm{BH}} (T_{\rm{BH}},z) = \frac{4 f^{e\pm}_{1/2} f^{e}_{eff} (E_e,z) + 2 f^\gamma_1 f^{\gamma}_{eff} (E_\gamma,z)}{4 f^{e\pm}_{1/2} + 2 f^\gamma_1}. 
\label{equ:feff_BH}
\end{equation}

\begin{figure}
\centering
\begin{tabular}{cc}
\includegraphics[width=0.47\columnwidth]{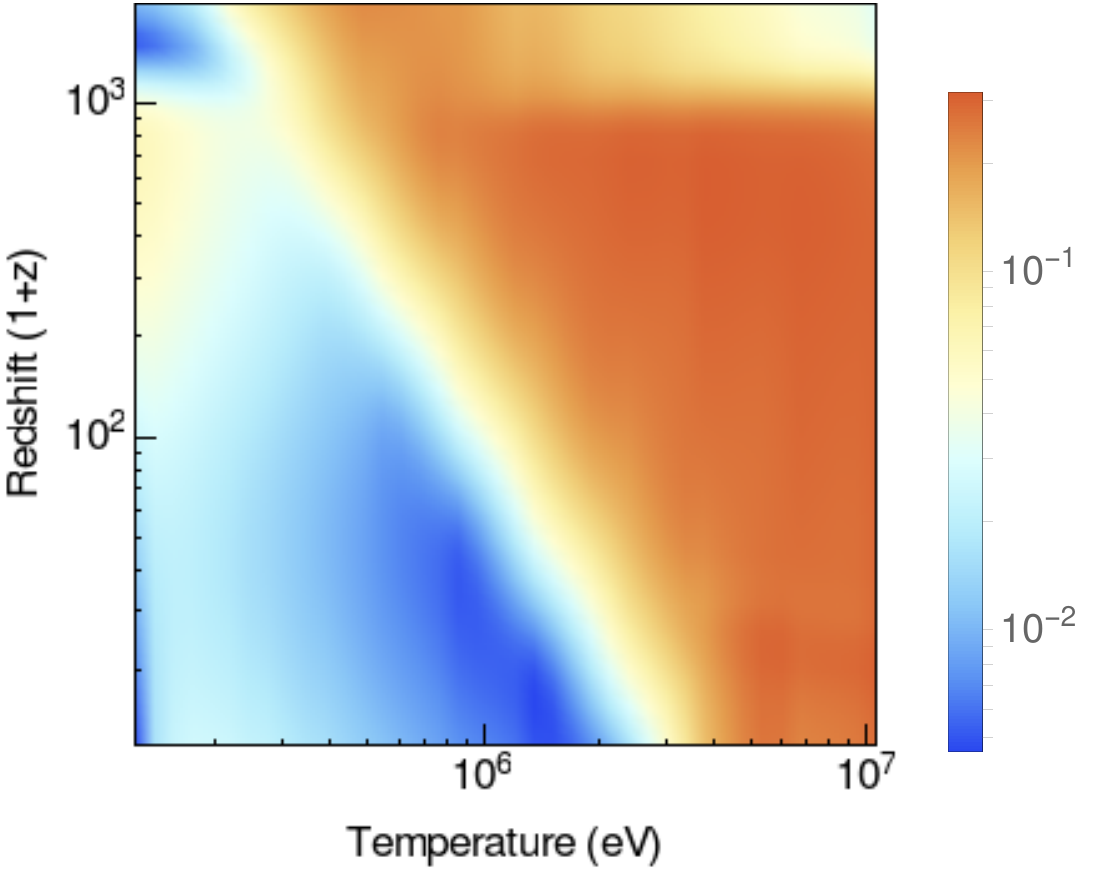} & \includegraphics[width=0.47\columnwidth]{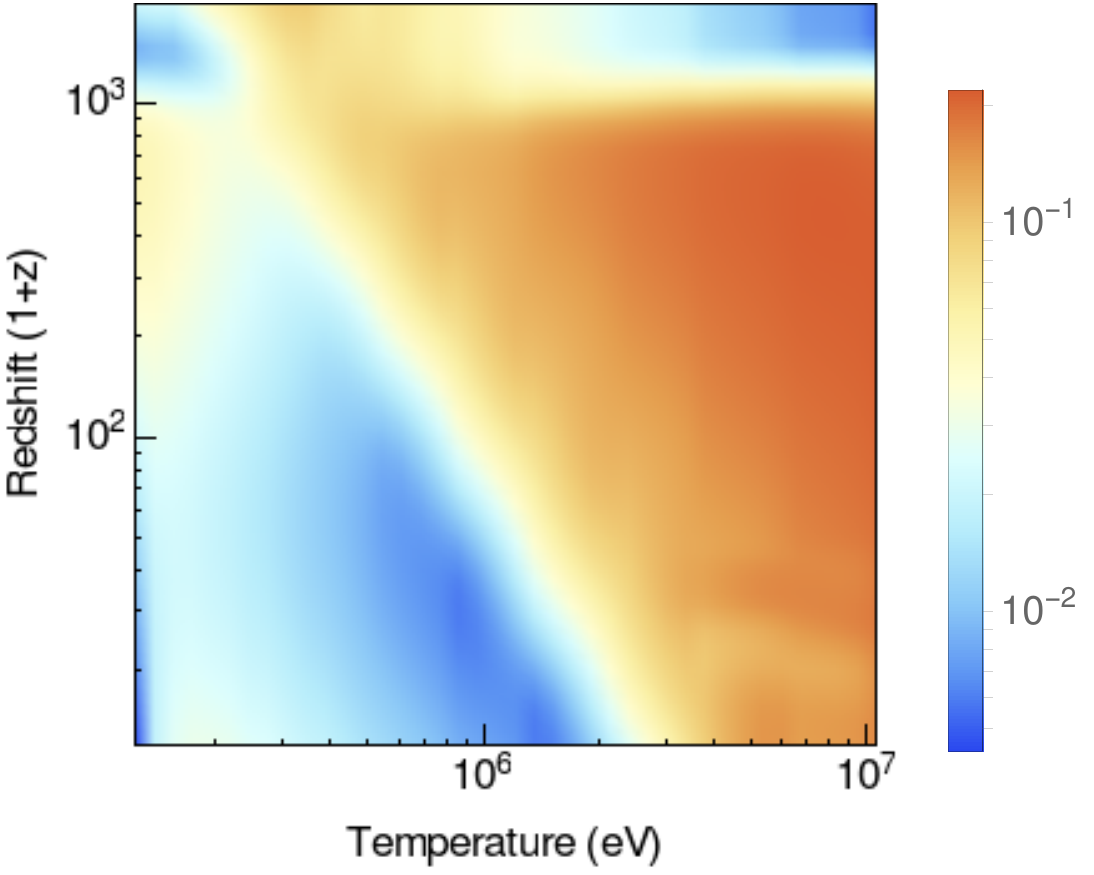} \\
\includegraphics[width=0.47\columnwidth]{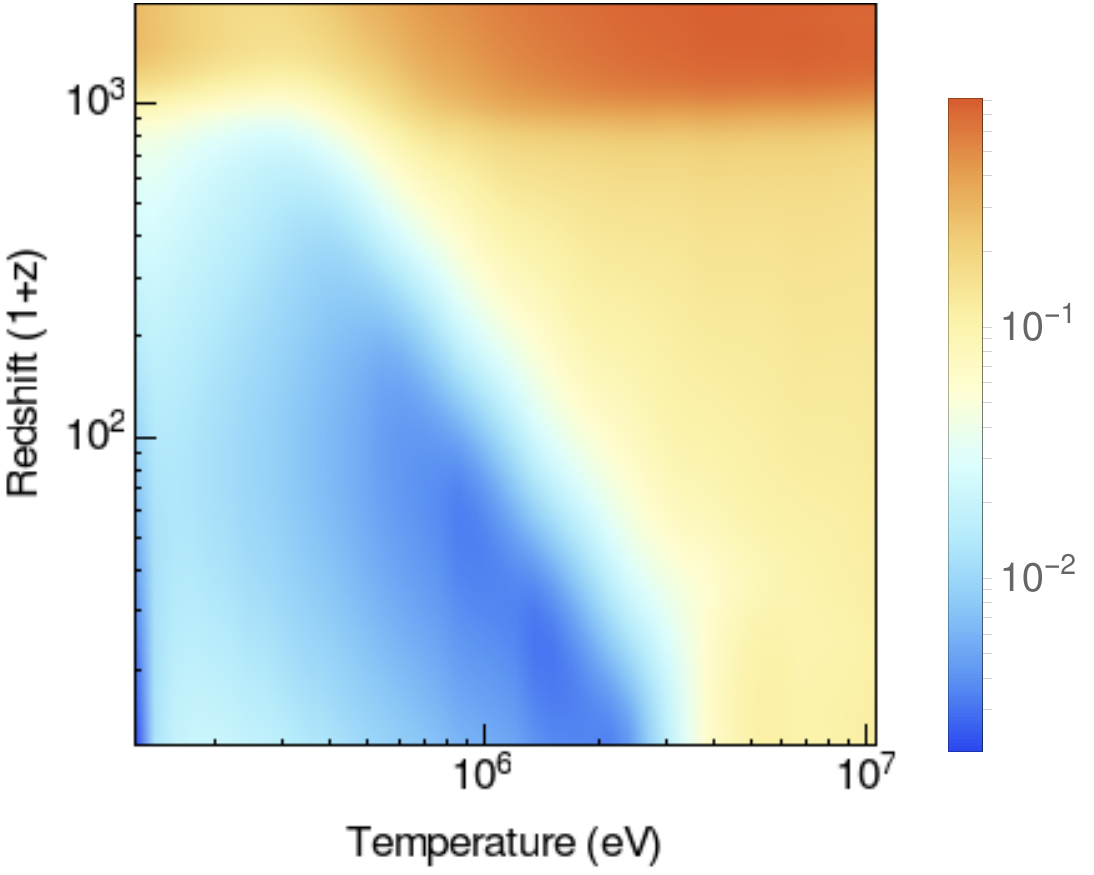} &
\includegraphics[width=0.47\columnwidth]{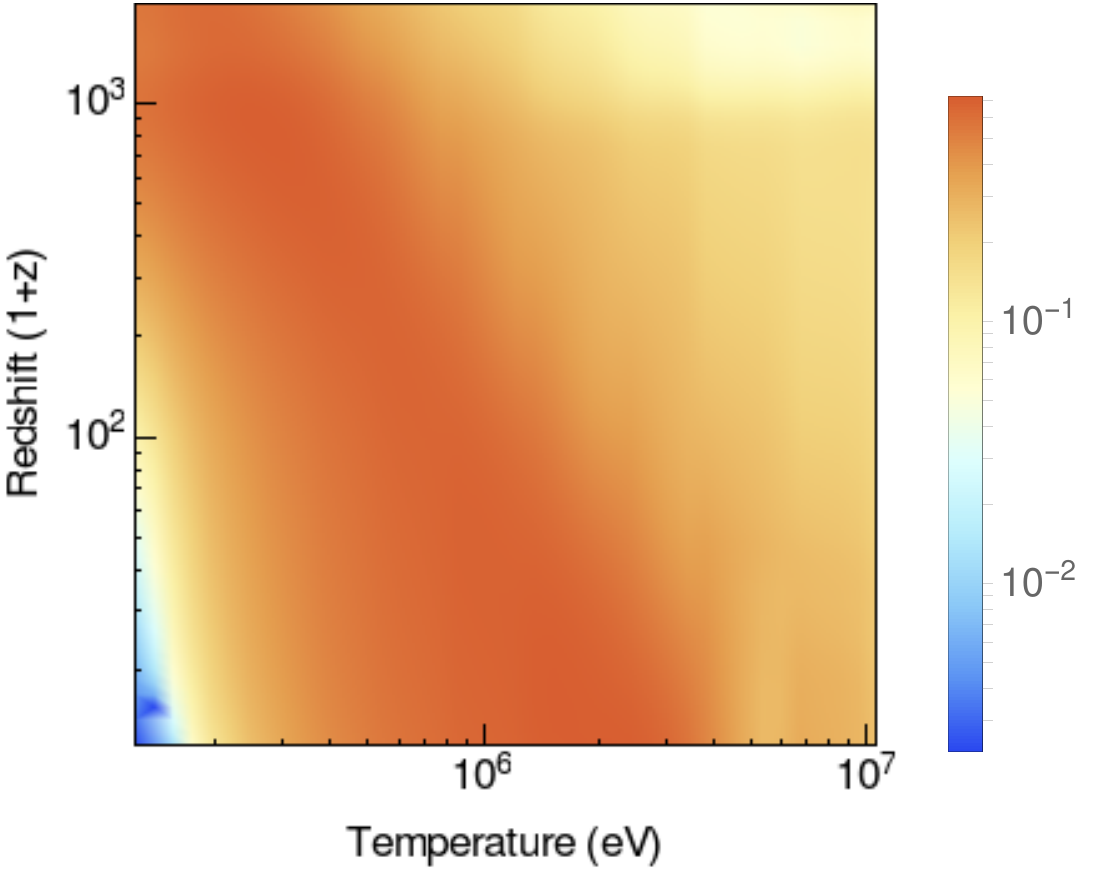} \\
\end{tabular}
\caption{Effective efficiencies of energy deposition from black holes that release electrons and photons in the manner described in Section \ref{blackholes}. From left to right, top to bottom, the curves are - Hydrogen Ionization, Lyman-Alpha Excitations, Heating, and Continuum Photons}
\label{fig:Effective_Efficiency}
\end{figure}

Using the decay efficiencies from Refs.~\cite{Slatyer:2015kla,Liu:2016cnk}, Figure \ref{fig:Effective_Efficiency} shows the effective efficiencies for PBH Hawking radiation for the mass range considered in this work. It should be noted that there appears a location where the efficiency drops drastically for all channels used in altering the ionization history, and this location shifts to later times for increasing temperatures. This drop will have a direct impact on the constraints for a given PBH mass.

\section{Results} \label{results}
To model the energy injection from PBHs and its impact on the recombination history and CMB, we utilize the CAMB~\cite{Lewis:1999bs,Howlett:2012mh} and HYREC~\cite{AliHaimoud:2010dx} codes. We incorporate two new parameters into the code, the PBH mass, $m_{\rm{BH}}$, and the ratio of the PBH density to the total dark matter density, $\Omega_{\rm{BH}}/\Omega_{\rm{DM}}$. With this modification to the recombination history, we fit CMB data using the COSMOMC code~\cite{Lewis:2002ah,Lewis:2013hha}. In Section \ref{recomb_hist} we determine how PBH energy injections alter the ionization history, in Section~\ref{CMB_alter} we determine the general effects the energy injection has on CMB anisotropies, and in Section \ref{black_hole_constr} we present constraints on the PBH density. 

\subsection{Recombination History} \label{recomb_hist}
\begin{figure}
\centering
\begin{tabular}{cc}
\includegraphics[width=0.47\columnwidth]{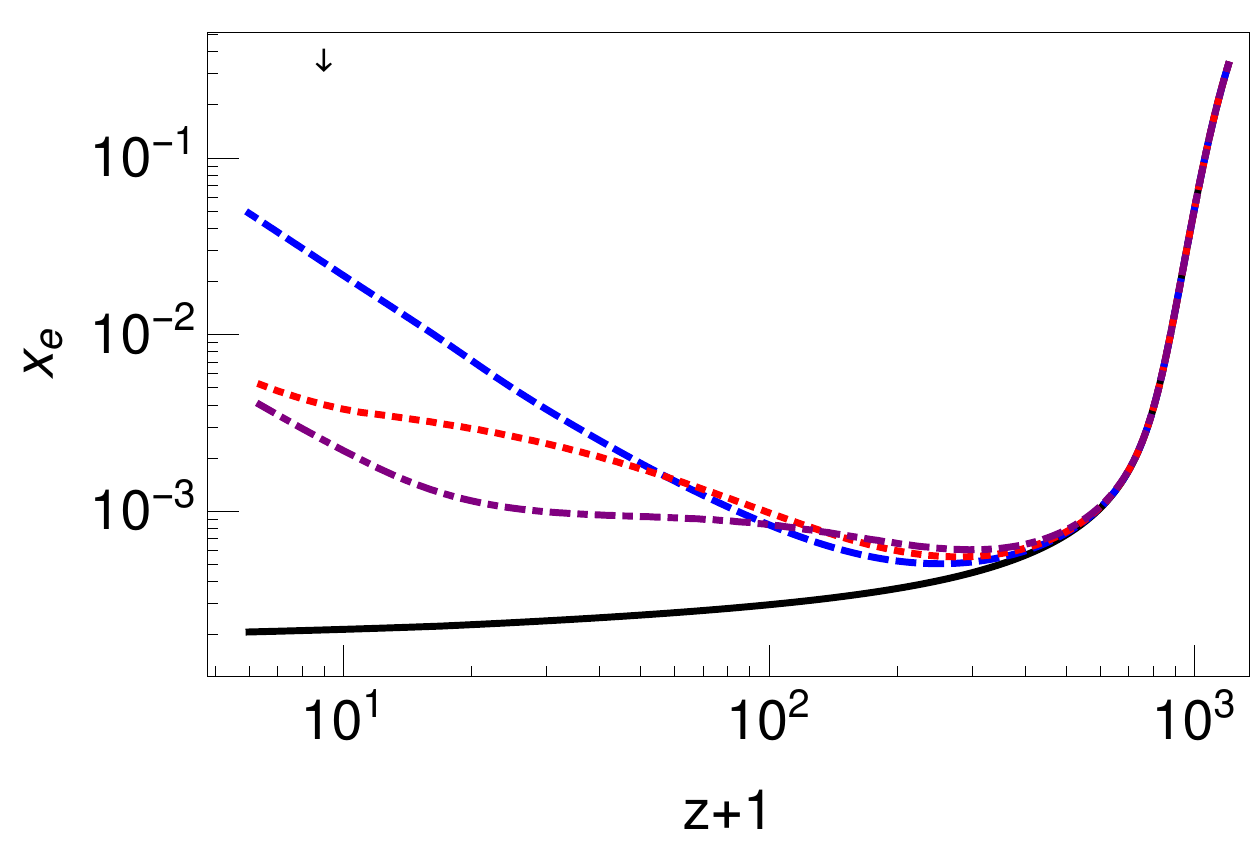} & \includegraphics[width=0.47\columnwidth]{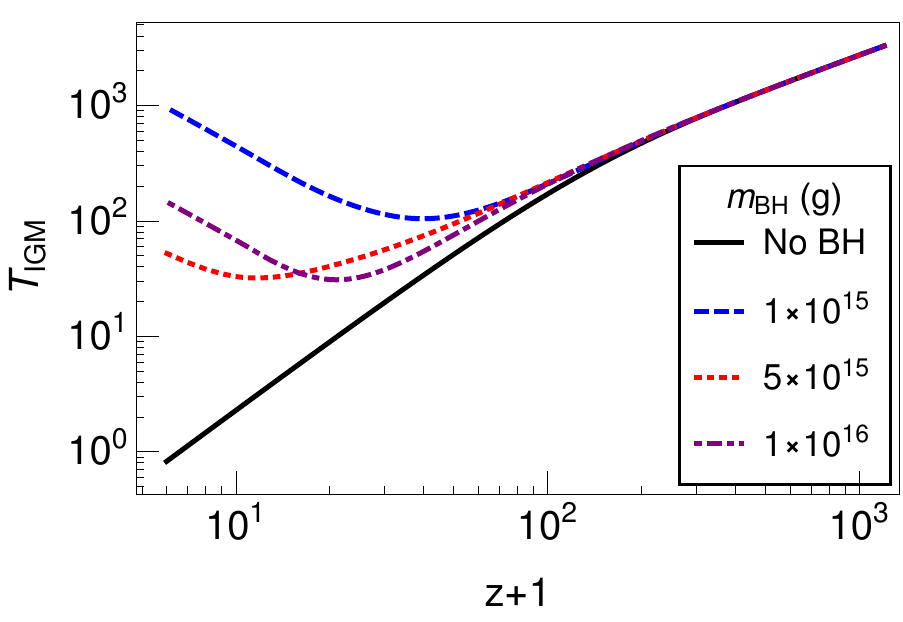} \\
\includegraphics[width=0.47\columnwidth]{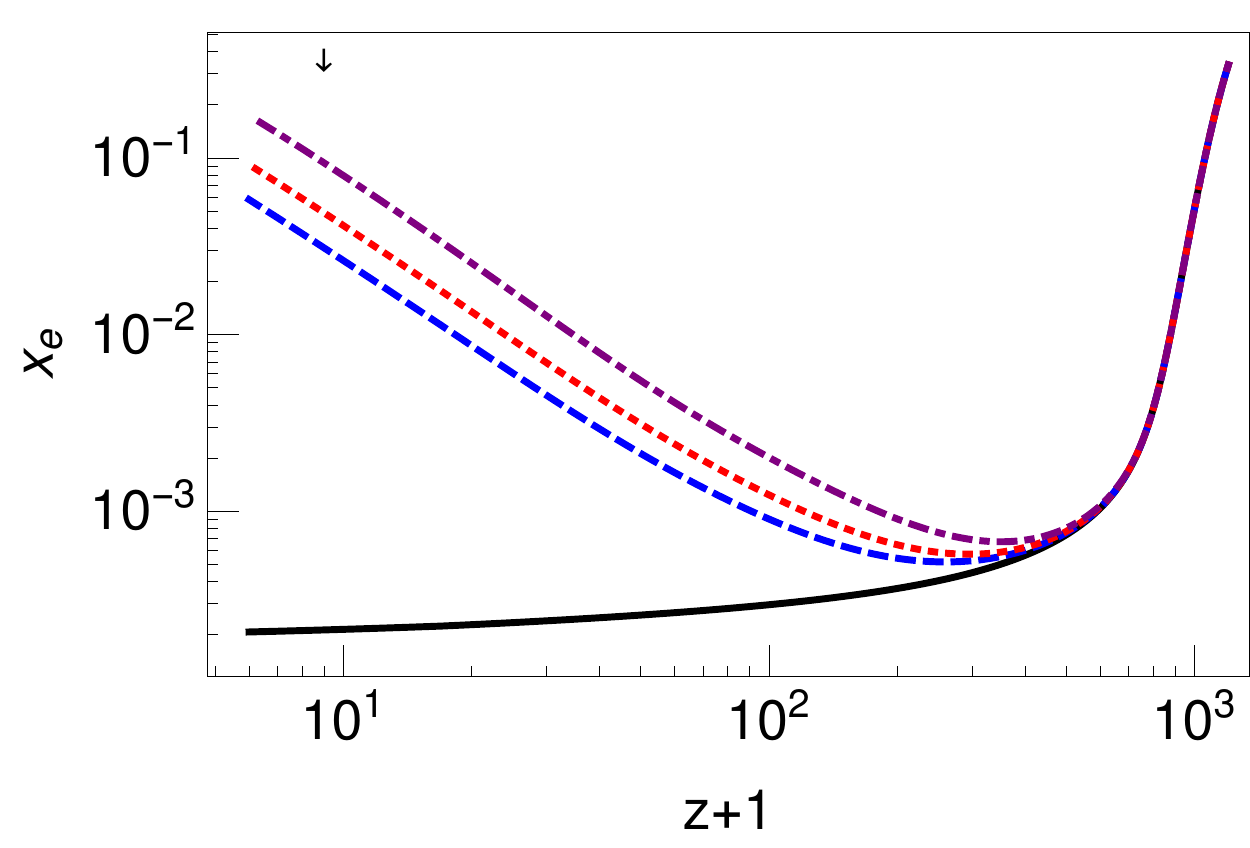} & \includegraphics[width=0.47\columnwidth]{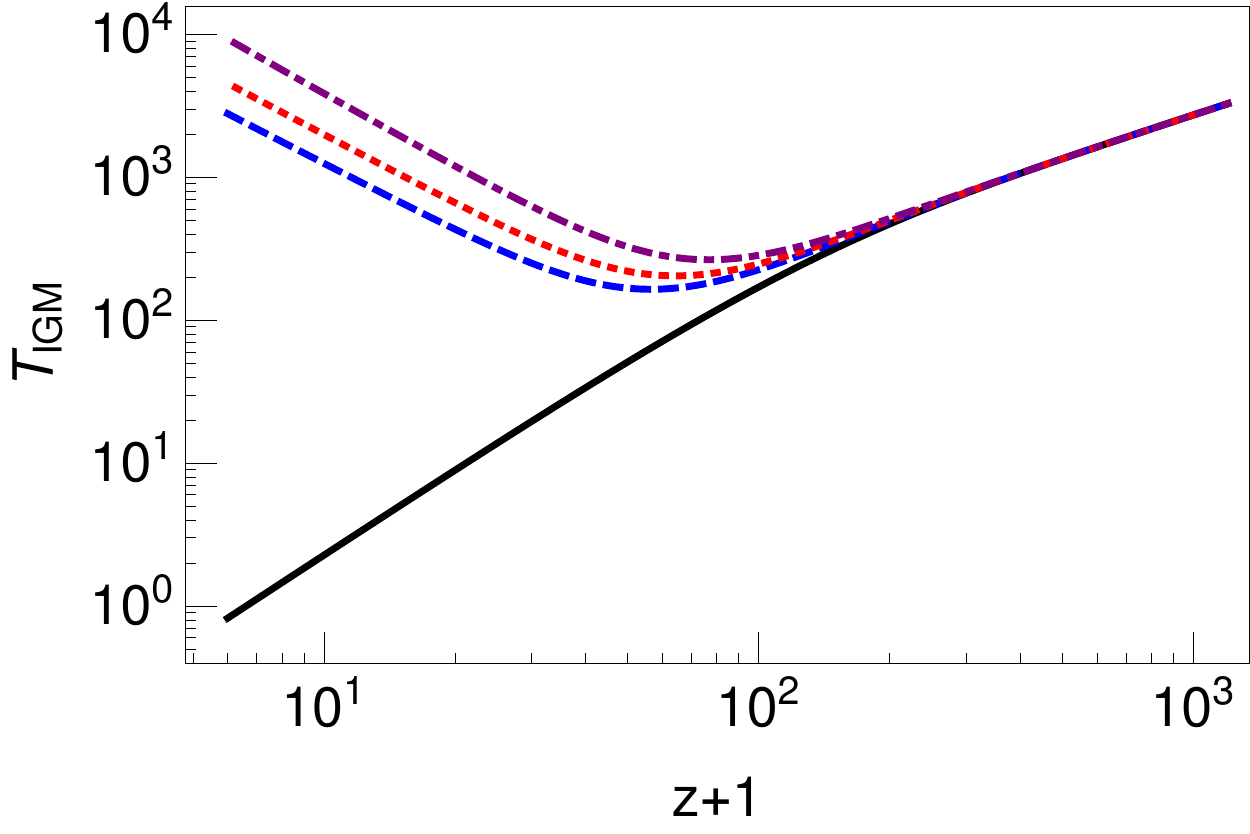} \\
\includegraphics[width=0.47\columnwidth]{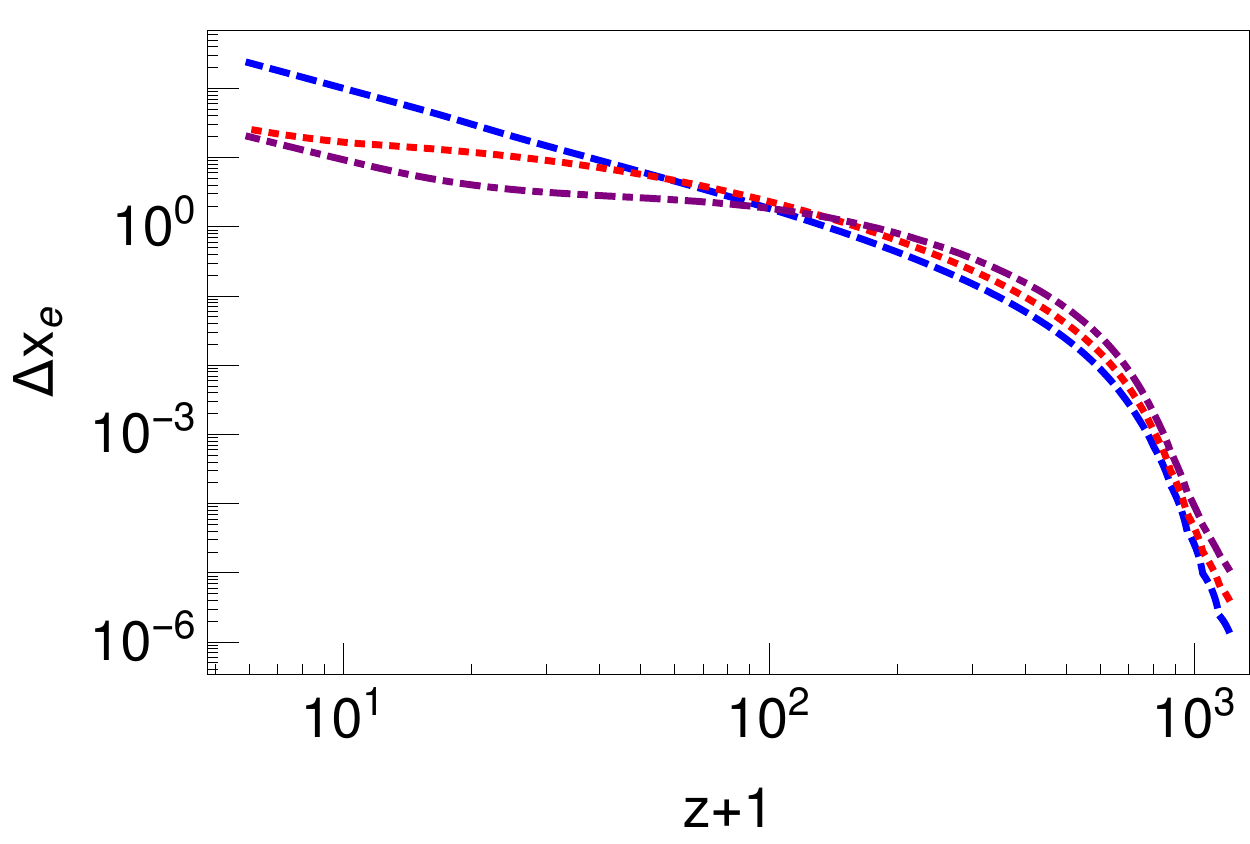} & \includegraphics[width=0.47\columnwidth]{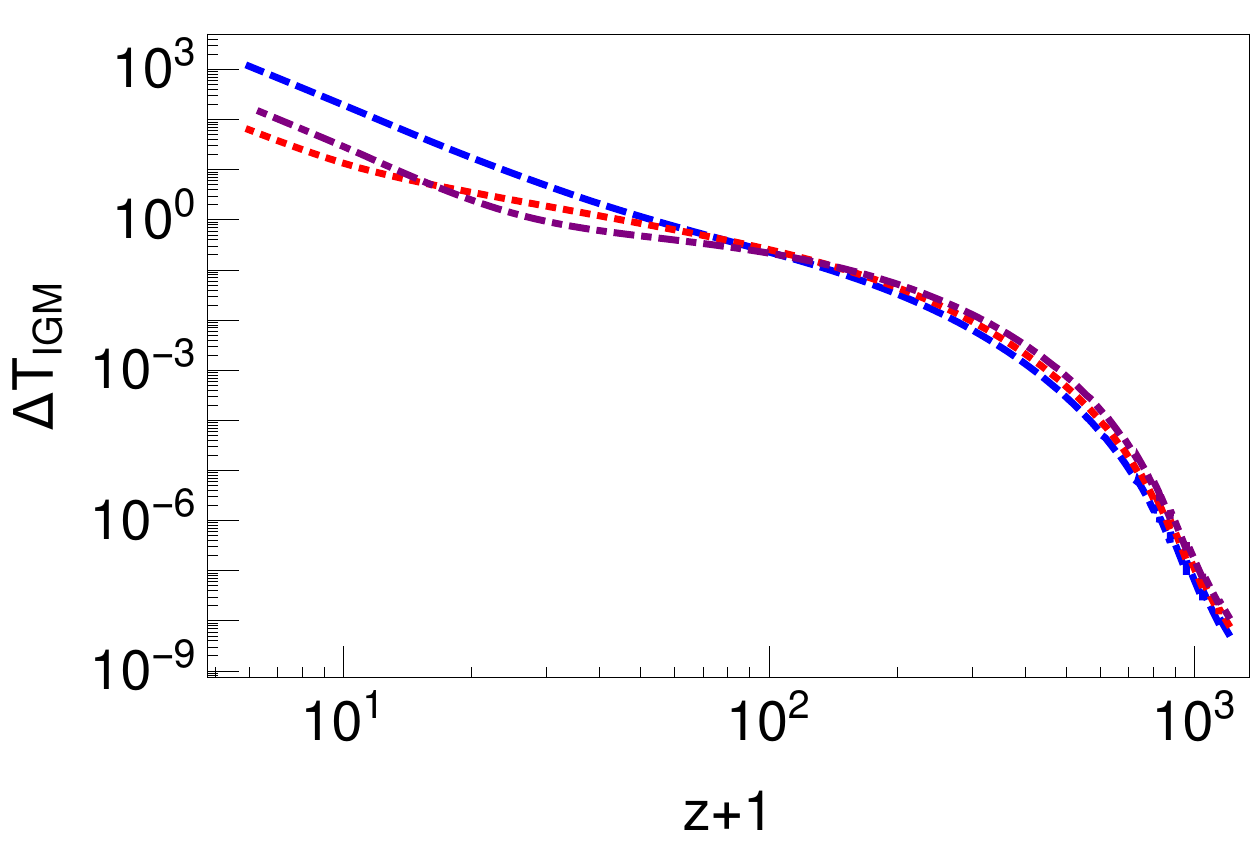} \\
\end{tabular}
\caption{Ionization fraction (left) and temperature of the IGM (right) due to PBH energy injection using effective efficiencies (top), using the ``SSCK'' prescription (middle), and percent change from the no PBH case (bottom). Bounds on the parameters are also plotted. The PBH density $\Omega_{\rm{BH}}$ for each mass is taken to be at the 95$\%$ confidence limit discussed in this work. The legend applies to all graphs.}

\label{fig:recomb_history}
\end{figure}

The alterations to $x_e$ and $T_{\rm{IGM}}$ are shown for a few example PBH masses in Figure~\ref{fig:recomb_history}. The percent change relative to the standard model with no PBHs is indicated. Figure~\ref{fig:recomb_history} shows that there is only a minor variation relative to the no PBH case in both $x_e$ and $T_{\rm{IGM}}$ at large redshifts, with the significant deviations only apparent at $\sim z=200-300$. The minimal deviation at high redshift supports the use of the effective efficiencies~\cite{Slatyer:2015jla}. For standard values of the cosmological parameters, $x_e$ and $T_{\rm{IGM}}$ are both well below the observational limits on these quantities~\cite{Bolton:2011ck,Bolton:2010gr}, 
\begin{equation}
\renewcommand{\arraystretch}{1.3}
\begin{tabular}{ccc}
$x_e(z \sim 7) = 0.66$ & $\quad$ & $x_e(z \sim 8) = 0.35$\\
$\log_{10}(T_{\rm{IGM}}(z=4.8))=3.9\pm0.1$ & $\quad$ & $\log_{10}(T_{\rm{IGM}}(z=6.08))=4.21^{+0.06}_{-0.07}$. 
\end{tabular}
\label{equ:recomb_bounds}
\end{equation}
We note that contributions from reionization and structure formation are not included in the calculation, though it may be possible that PBHs around the mass that we study can represent a significant contribution to reionization~\cite{Belotsky:2014twa}. 

Since there is a large variation in the ionization fraction at later times, assumptions made about the effective efficiencies weaken. In order to observe possible errors introduced due to this large deviation, an ionization history developed using the ``SSCK'' prescription~\cite{Madhavacheril:2013cna,Slatyer:2015jla} and the same cosmological parameters is also given in Figure \ref{fig:recomb_history}. While there is again a large variation at late times, up to two orders of magnitude, $x_e$ and $T_{\rm{IGM}}$ are still far below the observational constraints. Additionally during the period of interest at high redshift, they have only minor variation relative to the standard no PBH case.

\subsection{CMB Anisotropies} \label{CMB_alter}

%The CMB is the surface of last scattering for photons. The signal that is measured today is a result of their interactions with the IGM and is dependenton the ionization history. Different interactions with the IGM combine together to produce multiple correlation relationships. This work focuses on using anisotropy correlations to place constraints on PBH properties.

\begin{figure}
\centering
\begin{tabular}{ccc}
\includegraphics[width=0.30\columnwidth]{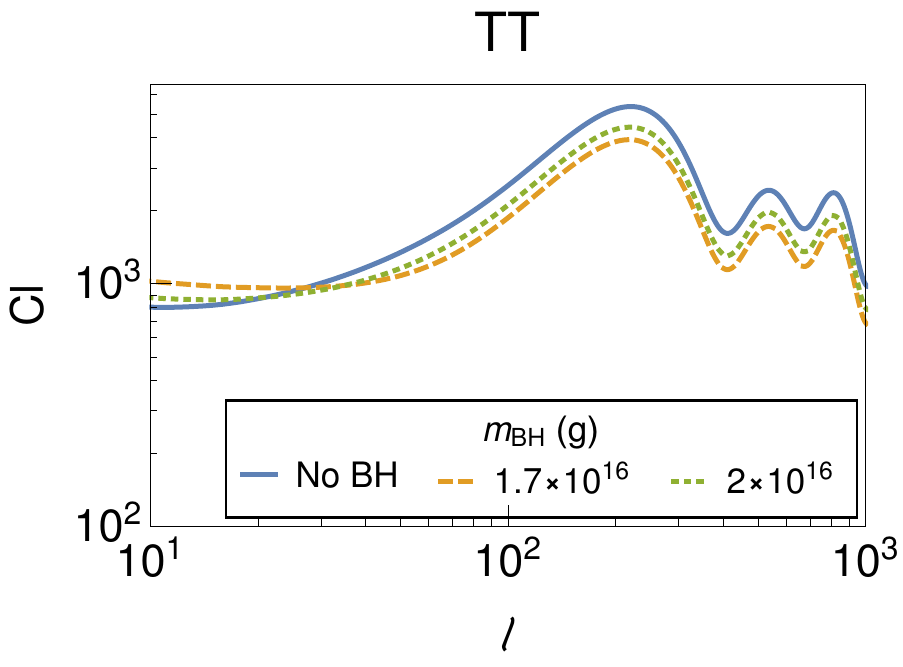} & \includegraphics[width=0.30\columnwidth]{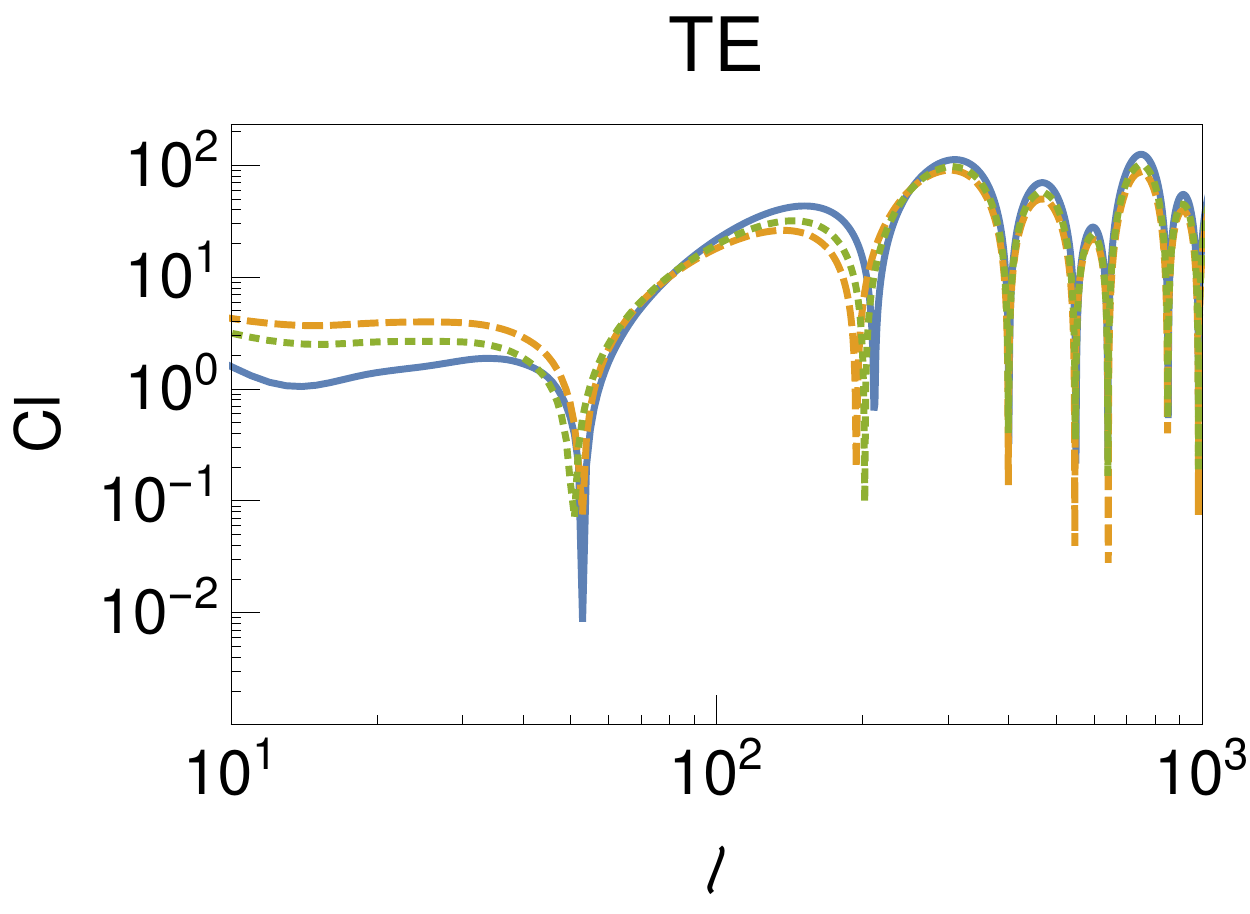} & \includegraphics[width=0.30\columnwidth]{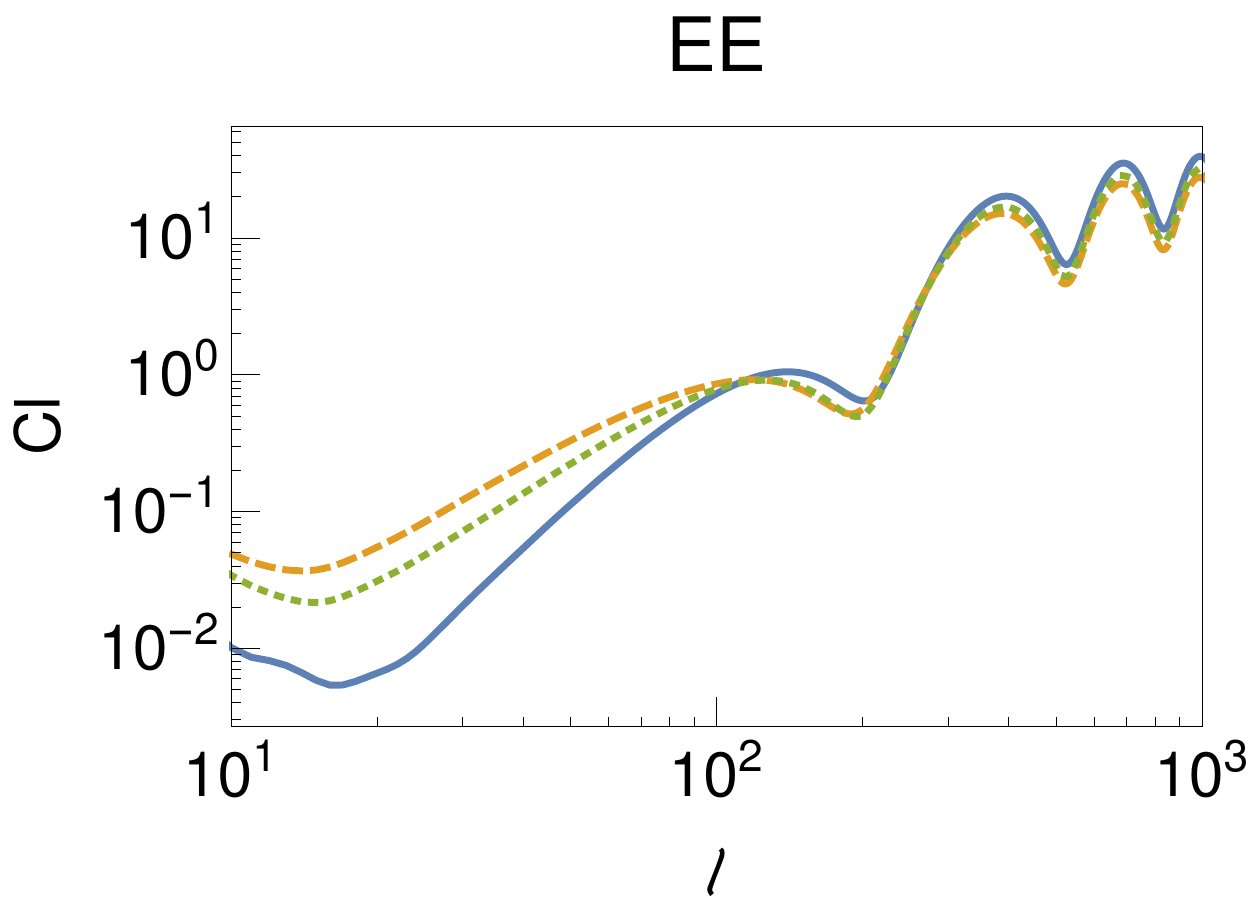} \\
\end{tabular}
\caption{Effect that various PBH masses have on the CMB for TT (left), TE (middle), and EE (right) correlations. The legend present in the TT panel also applies to the TE and EE panels.}
\label{fig:BlackHole_effect}
\end{figure}

PBHs also affect the measured CMB anisotropy power spectrum. Here we discuss the affect on the Temperature-Temperature (TT), Temperature-Polarization (TE), and Electric Polarization-Polarization (EE) power spectra. In Figure \ref{fig:BlackHole_effect}, TT, TE, and EE correlations are given for example PBH masses, assuming $\Omega_{\rm{BH}}$ composes all of the dark matter. As can be expected, larger changes arise from smaller mass PBHs due to both their larger radiation rate as well as their higher number density at equal mass densities.

The energy injection results in a scale-dependent deviation from the standard case with no PBHs; there is an increase in the power spectrum at small multipoles and a decrease at large multipoles. This behavior can be understood by noting that the width of the last scattering surface increases because of the PBH energy injection. Perturbations on scales smaller than the width of the last scattering surface are suppressed, as can be seen most easily seen for TT correlations. In addition the TE and EE spectra shift with an energy injection. These shifts are due to monopole perturbations to the quadrapole polarization that are introduced with the increased width of the scattering surface~\cite{Padmanabhan:2005es}. 

\subsection{PBH constraints} \label{black_hole_constr}

\begin{figure}
\centering
\begin{tabular}{rrr}
\includegraphics[width=0.30\columnwidth]{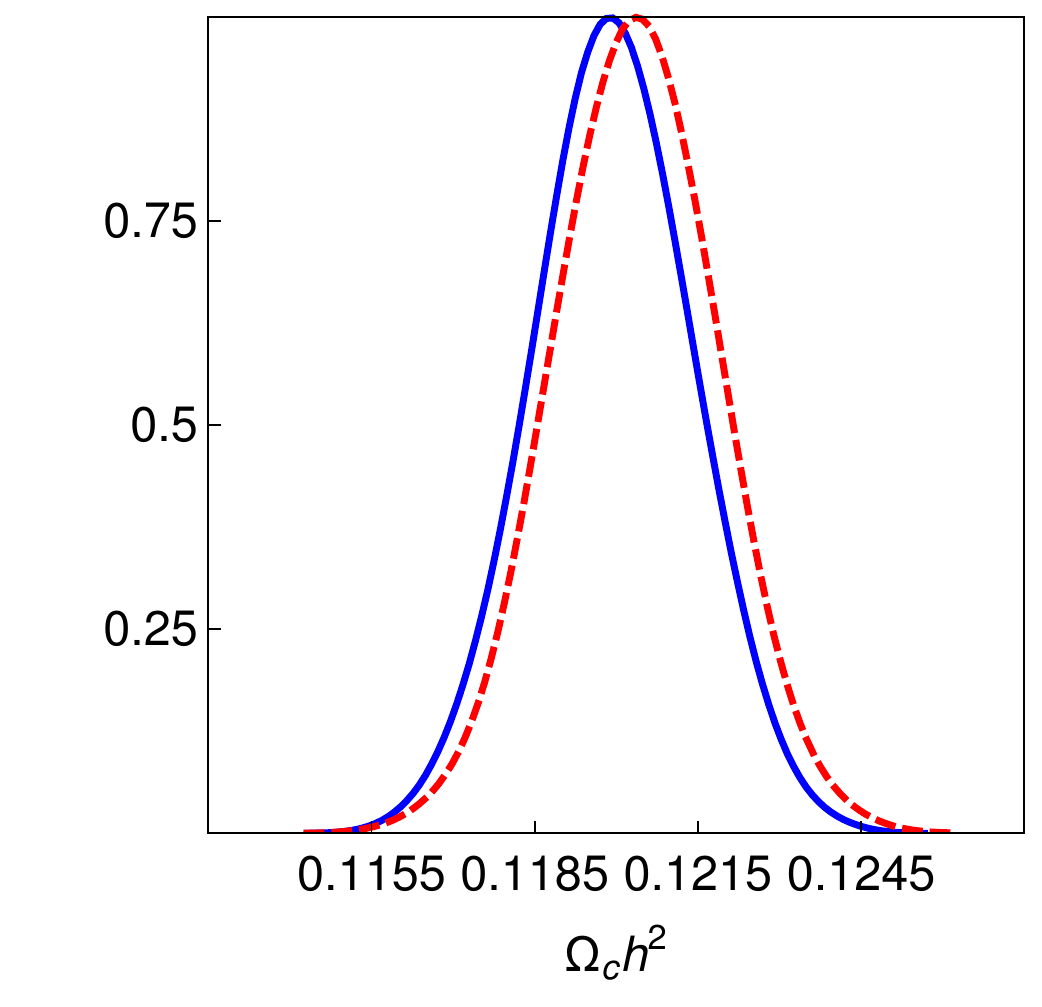} & \includegraphics[width=0.30\columnwidth]{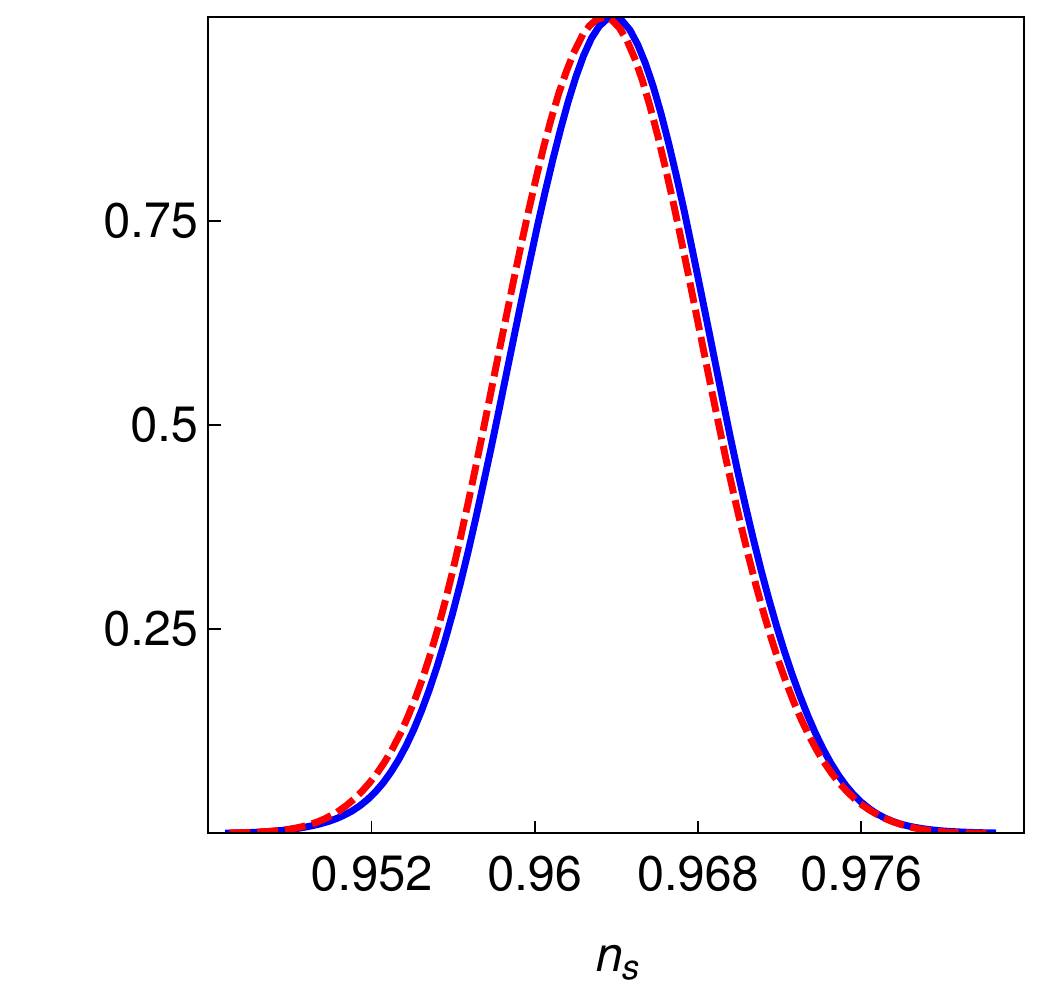} & \includegraphics[width=0.30\columnwidth]{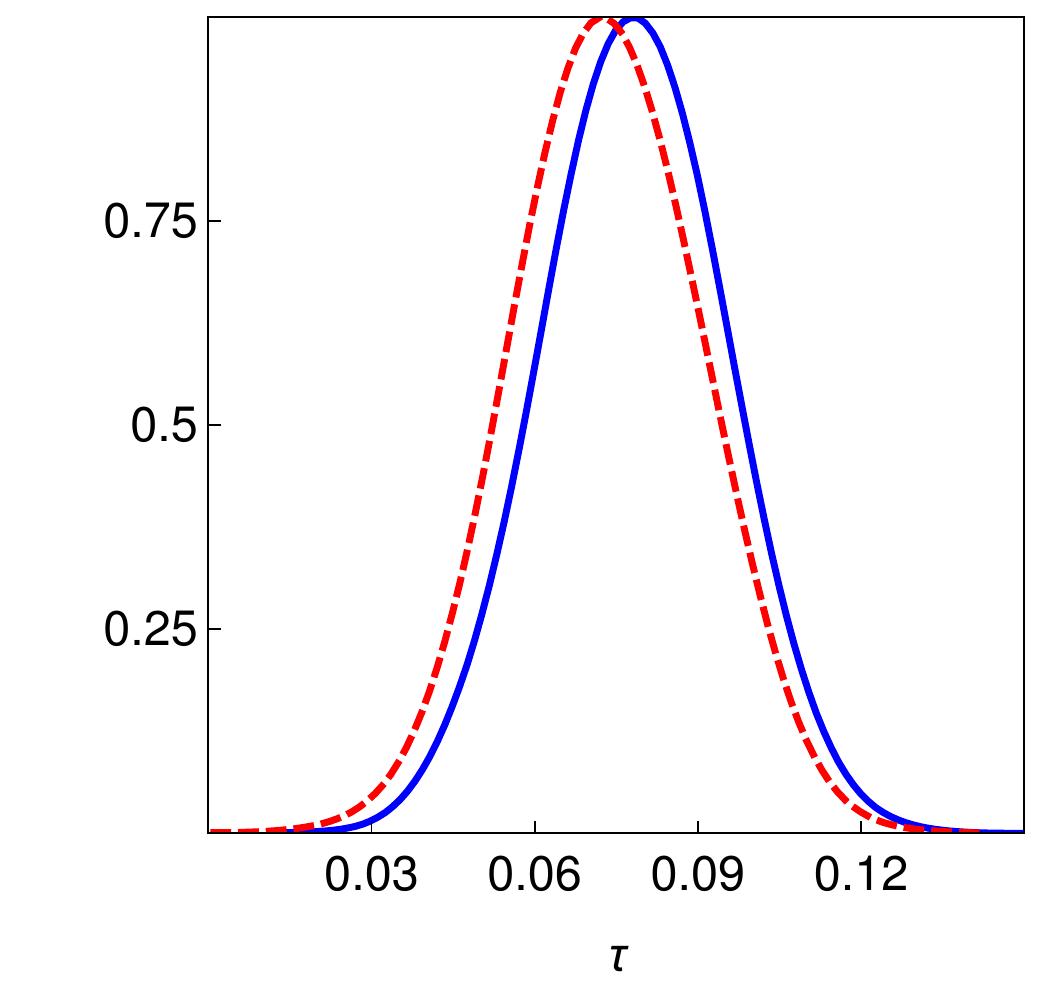} \\
\includegraphics[width=0.30\columnwidth]{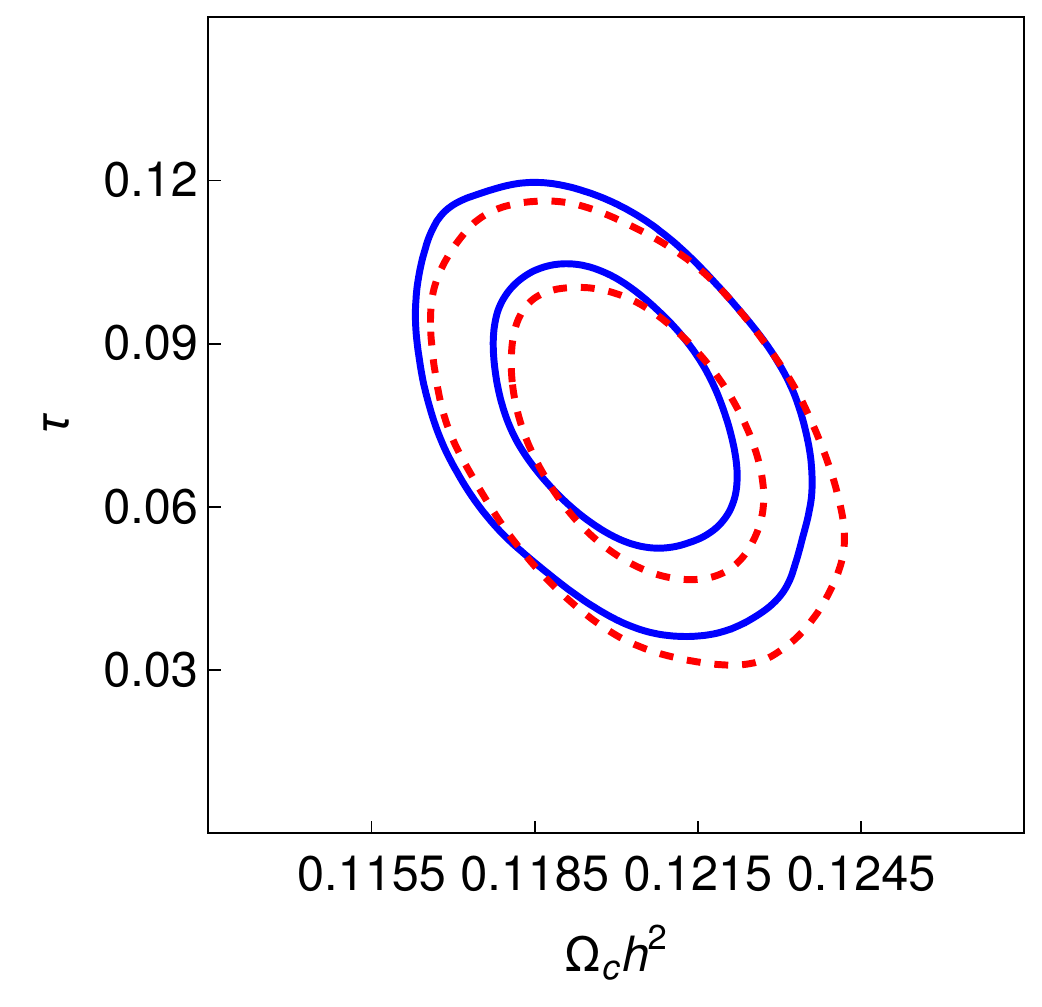} & \includegraphics[width=0.30\columnwidth]{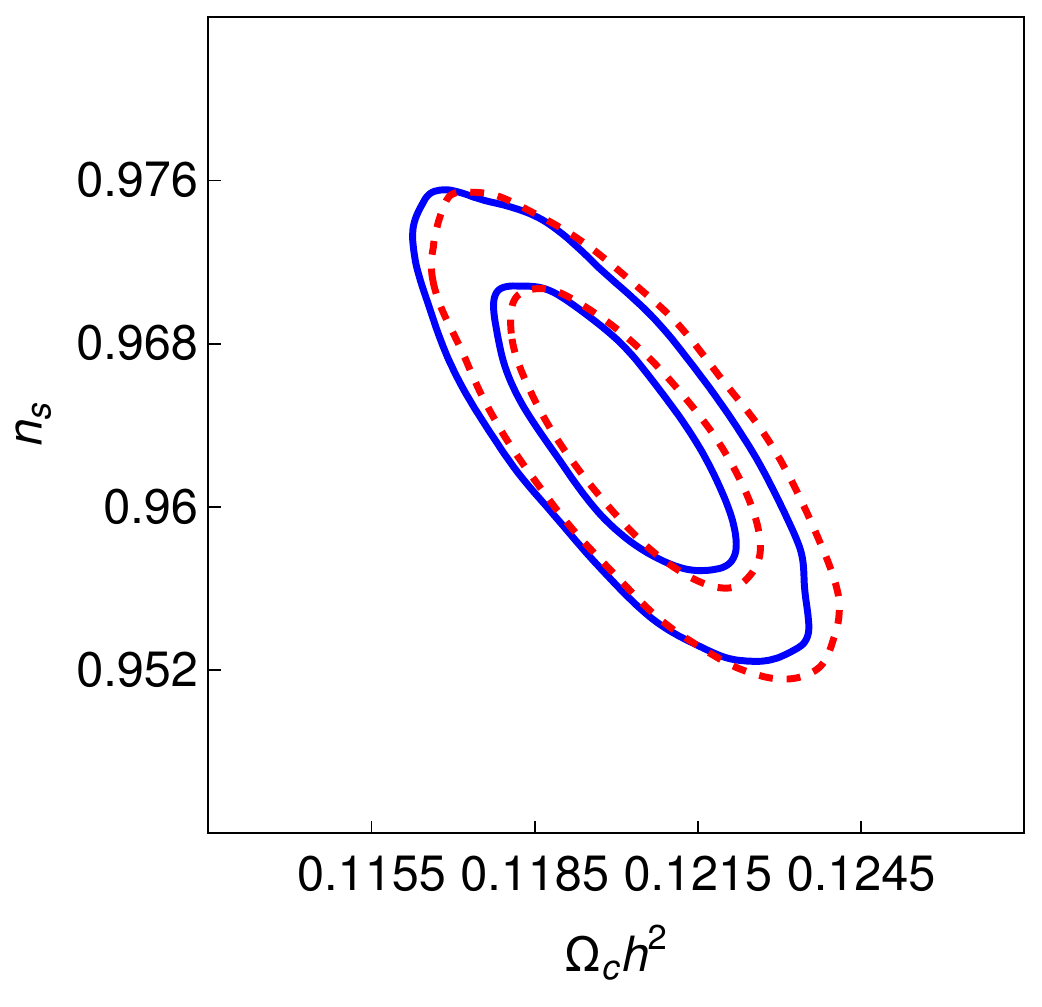} & \includegraphics[width=0.30\columnwidth]{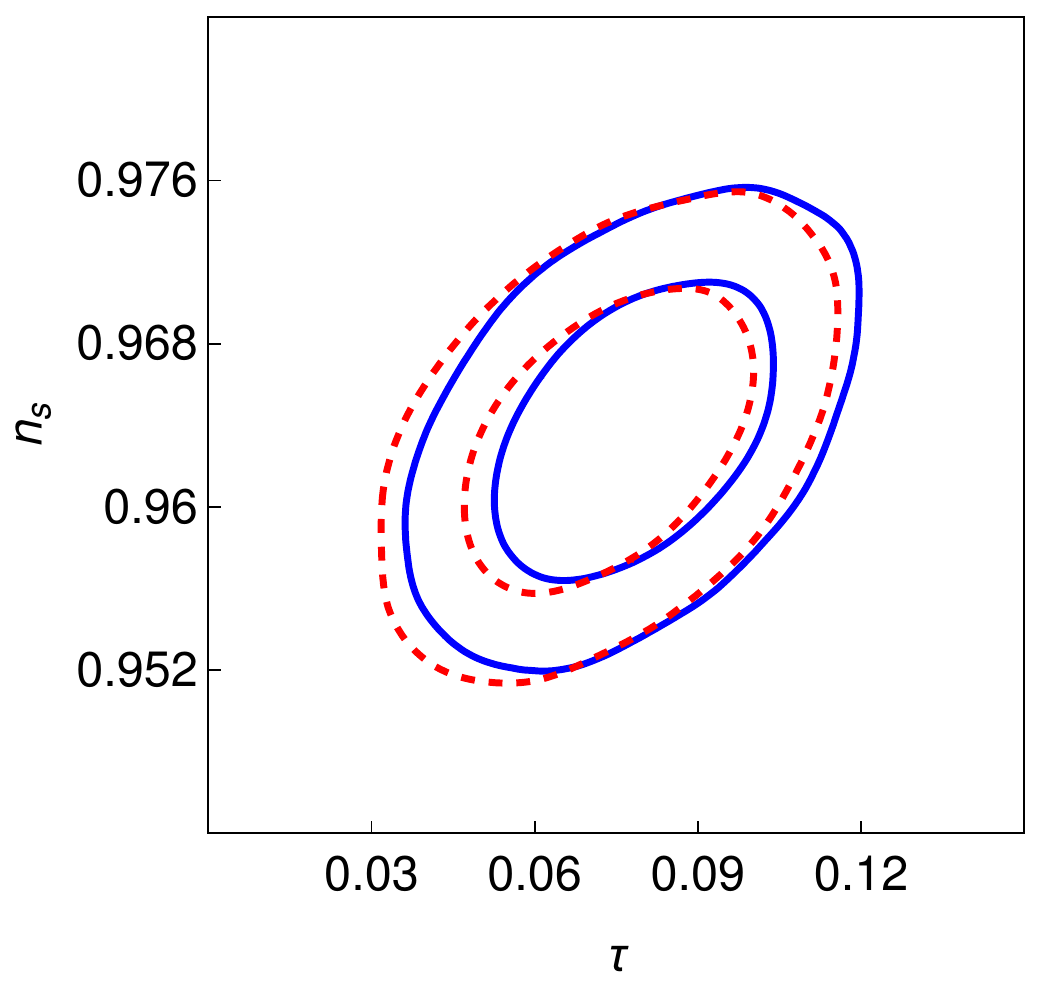}
\end{tabular}
\caption{Change in the posterior probability distributions for a few principal cosmological parameters for PBHs of mass $2\times10^{16}g$. The optical depth is $\tau$ and the spectral index is $n_s$. {\it Top row}: Single variable distributions with most probable values normalized to one, where blue (solid) lines represent no PBH case while red (dashed) includes PBHs. {\it Bottom row}: Correlations between different parameters, with inner and outer curves correspond to 68\% and 95\% confidence levels respectfully. 
}
\label{fig:New_Param_Limits}
\end{figure}

To place constraints on the abundance of PBHs, we fit the simulated spectra to Planck half mission data~\cite{Aghanim:2015xee}. The likelihood used was TT, TE, EE+lowP for $l \geq 30$ and a Temperature-Electric Polarization-Magnetic Polarization correlation for $l \leq 29$. Figure~\ref{fig:New_Param_Limits} shows the posterior probability densities for a few example cosmological parameters, both with and without PBHs.  As can be seen, the distributions are consistent with each other, up to minor shifts well within experimental uncertainties. Note that this has been highlighted in previous studies of the impact of dark matter annihilation and decay on the CMB~\cite{Slatyer:2016qyl}. 

\begin{figure}
\centering
\includegraphics[width=0.84\columnwidth]{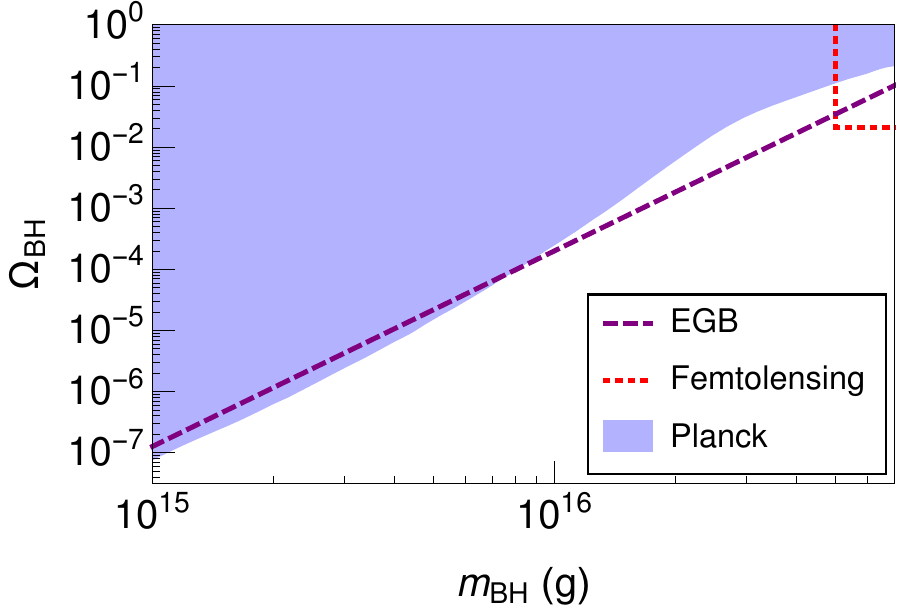}
\caption{CMB exclusion bounds for $\Omega_{\rm{BH}}$ at the 95\% confidence level (shaded region) compared with the same exclusion bound enforced by EGB, assuming 100\%~of the background produced by PBHs (long dashes). Also included is an estimation for the bound that is imposed due to femtolensing (short dashes) considered in~\cite{Carr:2016drx}.}
\label{fig:EGB_Comparison}
\end{figure}

Since there is little variation in the base cosmological parameters, for computational convenience to set upper limits on $\Omega_{\rm{BH}}$ we take the six principle cosmological parameters to be fixed at their best fit values in the case of no additional energy injection~\cite{PlanckTable:2015jan}. Figure~\ref{fig:EGB_Comparison} shows the result of the 95\% confidence limit, where the confidence limit is defined as the cumulative distribution centered around the median, which corresponds closely to the peak of the distributions in~Figure~\ref{fig:EGB_Comparison}. The constraint follows the expected inverse cube relationship to the PBH mass which is predicted by the energy injection formula.

In addition to the cubic dependence on mass, there is also a highly nonlinear relationship to the effective efficiency. This nonlinearity is most prevalent at a PBH mass around $1-4 \times 10^{16}$ g. Comparing effective efficiencies, in Figure \ref{fig:Effective_Efficiency}, the trend is correlated with the efficiency values that occur near the time of recombination. As the efficiency value decreases, it is required for a larger amount of total energy to be created in order to produce the same effect. For this reason, as the efficiency experiences a large decrease, the allowable maximum mass fraction increases.

We note that since the six base cosmological parameters were fixed, the PBH abundance may be more strongly constrained than in a model in which more parameters are allowed to vary. To check this, we compare to the case in which the base cosmological parameters are allowed to vary for a single PBH mass of $2 \times 10^{16}$ g. We find that by freeing all of the cosmological parameters, the constraint on $\Omega_{\rm{BH}}$ may be weakened by up to a factor of three. However, as stated above for computational convenience we have decided to fix the base cosmological parameters for our main bounds on $\Omega_{BH}$. 

As indicated above previous analyses have used the EGB to constrain PBHs in the mass regime $10^{15}-10^{17}$ g~\cite{Poulin:2016anj}. Since PBHs emit a mass-dependent gamma-ray spectrum, there is an upper bound on their density before they would be excluded by EGB measurements. Following the prescription outlined in Ref.~\cite{Carr:2009jm}, the number density of photons $n_{\gamma 0}$ with energy $E_{\gamma 0}$ and their intensity is
\begin{equation}
n_{\gamma 0}(E_\gamma) = \frac{\Gamma_{\rm{BH}}}{M_{\rm{BH}}}E_\gamma \int_{t_{\rm{min}}}^{\rm{min}(t_0,\tau)}dt(1+z)^{-2}\dfrac{\dot{N}_\gamma}{E_\gamma}(M_{\rm{BH}},(1+z)E_\gamma)
\end{equation}
\begin{equation}
I=\frac{c}{4\pi}n_{\gamma 0}
\end{equation}
where $t_{\rm{min}}$ is the time when photon creation begins. The quantity $\dot{N}_\gamma/E_\gamma(M_{\rm{BH}},E_\gamma)$ is the photon spectrum given by Equation (\ref{equ:Hawking-Dist}), which we take at the high energy limit. For PBHs in the mass range studied, peak intensity occurs at $\sim1-30$ MeV. Constraints were derived by matching the intensity to the upper bound of the COMPTEL EGB experimental data~\cite{Weidenspointner:1999}. 

EGB constraints are also shown in Figure \ref{fig:EGB_Comparison} as well as those imposed by femtolensing~\cite{Carr:2016drx}. We find that Planck provides the strongest constraint on the abundance of PBHs for masses $\sim 10^{15}-10^{16}$ g, while the EGB dominates for masses $\gtrsim 10^{16}$ g. Note that this conclusion differs from that of Ref.~\cite{Poulin:2016anj}. The Planck constraint deviates from a linear relation because of the model for effective efficiencies. 

\section{Conclusions} \label{conclusions}

PBHs are of great interest in cosmology. They reveal conditions in the early universe and can serve as a dark matter candidate. There are several standard mechanisms that have been proposed to detect PBHs; these include detection of Hawking radiation, detection of radiation produced from accretion disks, and gravitational lensing. Each method is capable of targeting different PBH mass ranges. In this paper, we have focused on PBHs with masses in the range $10^{15}-10^{17}$ g. We have improved and made more precise the constraints in this mass range using the CMB and EGB. For our CMB bound, we model the energy absorption not as instantaneous, but rather using redshift dependent efficiency. The energy injection results in an increase in the ionization fraction at late times as well as an increase in the IGM temperature, leading to distortions of the CMB anisotropies. Larger fractional changes occur at large multipoles because of the increase of the width of the last scattering surface. 

Using Planck data, we show that CMB distortions from Hawking radiation allow for stringent constraints on the density of $10^{15}-10^{17}$ g PBHs of $\Omega_{\rm{BH}} \lesssim 3.3 \times 10^{-9} (m_{\rm{BH}}/M_\star)^{3.8}$. We show that for mass $\sim 10^{15}-10^{16}$ g, the CMB constraints are stronger than the constraints from the $\sim 1-30$ MeV EGB, which imply, 
%The EGB is another method that can be used for constraining PBH density in this mass range. However, with current experimental results, it is found to be less constraining, 
$\Omega_{\rm{BH}} \lesssim 1.4 \times 10^{-8} (m_{\rm{BH}}/M_\star)^{3.2}$. Constraints imposed by CMB spectral distortions from Hawking radiation producing sub 10.2 eV photons are also much weaker than our constraint.

In the future, our theoretical analysis may be improved by including a mass spectrum of PBHs. In addition, even though we have used the EGB to bound the contribution of PBHs, it may be interesting to consider the EGB as a signal of PBHs. This is an exciting possibility because the origin of this $\sim$ MeV gamma-ray background is not yet known~\cite{Strigari:2005hu,Ruiz-Lapuente:2015yua,Horiuchi:2010kq}. Future missions to measure MeV gamma-rays will be especially important for the study of PBHs~\cite{DeAngelis:2016slk}.

\section*{Acknowledgements} \label{acknowledgements}

The authors thank Daniel Meerburg for helpful discussion and modified HyREC codes. We also thank Tomohiro Harada and Tracy Slatyer for helpful discussions. BD acknowledges support from DOE Grant DE-FG02-13ER42020. LES acknowledges support from NSF grant PHY-1522717. YG thanks the Mitchell Institute for Fundamental Physics and Astronomy and Wayne State University for support. SC acknowledges support from NASA Astrophysics Theory grant NNX12AC71G. SW is supported in part by NASA Astrophysics Theory Grant NNH12ZDA001N and DOE grant DE-FG02-85ER40237.

\section*{Appendix: Continuum Photons} \label{appendix}

Continuum photons affect the CMB by creating spectral distortions~\cite{Slatyer:2015kla}. Ref.~\cite{Zavala:2009mi} investigated limits to these spectral distortions modeling the distortions as a Bose-Einstein distribution with a chemical potentioal $\mu$, $\mu$-type distortions. In order to get a baseline estimate on the effect of the continuum photons injected by PBH on the CMB a similar approach was taken. Assuming the injection will alter the perfect blackbody spectrum with the same $\mu$-type distortions, these distortions will be approximately

\begin{equation}
\mu = 1.4\frac{\delta \rho_\gamma}{\rho_\gamma} = 1.4\int_{t_1}^{t_2}  \frac{\dot{\rho}_\gamma}{\rho_\gamma} dt,
\label{equ:Spec_Dist_mu}
\end{equation}
where $\rho_\gamma$ and $\dot{\rho}_\gamma$ are the energy density of the CMB and the distortion injection rate on the CMB respectfully. $\dot{\rho}_\gamma$ becomes the injection rate directly from the black hole energy going into the continuum. The result is  

\begin{equation}
\mu = 1.4\int_{z_1}^{z_2}  \frac{(dE/dVdt)_{\rm{BH},\rm{Cont.}}}{\rho_c \Omega_\gamma (1+z)^4} \frac{dz}{(1+z) H(z)},
\label{equ:Spec_Dist_BH}
\end{equation}
with $\Omega_\gamma h^2 \sim 2.47 \times 10^{-5}$.

\begin{figure}
\centering
\includegraphics[width=0.84\columnwidth]{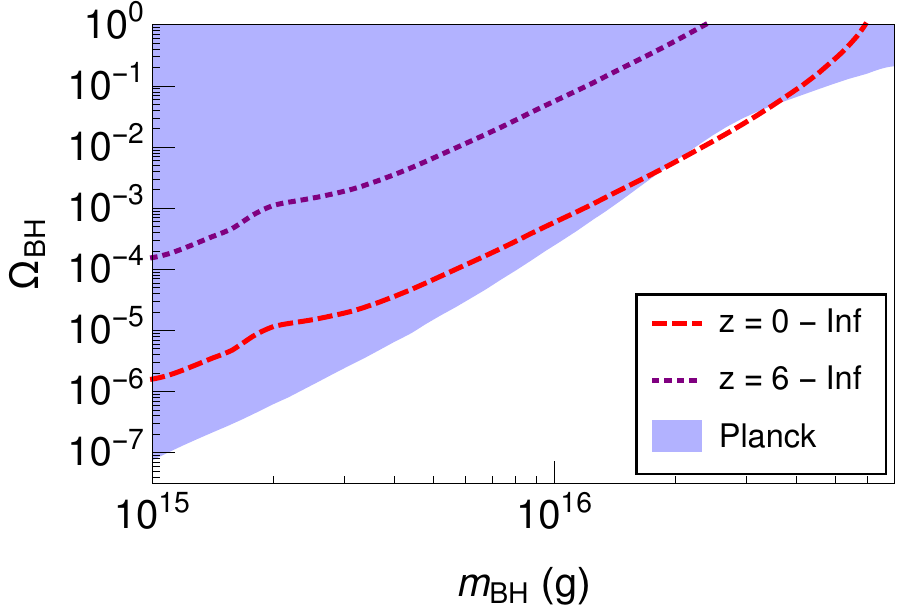}
\caption{Exclusion bounds for the fraction of dark matter that can be composed of black holes at the 95$\%$ confidence level compared with the same exclusion bound produced through spectral distortions. The various curves correspond to different integration limits in Equation (\ref{equ:Spec_Dist_mu}). The most constraining uses $z=0$ to infinity, and the second uses $z=6$ to infinity.}
\label{fig:Spectral_Distortion_Comparison}
\end{figure}

Current limits on these distortions are $|\mu| = 9.0 \times 10^{-5}$ at two sigma~\cite{Zavala:2009mi}. A comparison of this constraint and the CMB value discussed in this work is shown in Figure \ref{fig:Spectral_Distortion_Comparison}. As can be seen, the constraints produced by this work are much stronger, several orders of magnitude, than limits produced assuming a $\mu$-type or similar distortion. The only limits that approache the CMB result are those that consider alterations at extremely late times. These however can be ignored due to CMB photons at this time being much cooler than the 10.2 eV limit for these injections. Thus, the actual alterations at this time is further reduced.

\begin{thebibliography}{99}
%\cite{Hawking:1971ei}
\bibitem{bib:pbh} 
  S.~Hawking,
  %``Gravitationally collapsed objects of very low mass,''
  Mon.\ Not.\ Roy.\ Astron.\ Soc.\  {\bf 152}, 75 (1971).
  %%CITATION = MNRAA,152,75;%%
  %357 citations counted in INSPIRE as of 08 Dec 2016
  
%\bibitem{1966AZh....43..758Z}%
   Y.~B. Zel'dovich, ~I.~D. Novikov,
   Astron. Zh. volume 43, pages 758, 1966;
   
%\bibitem{1967SvA....10..602Z}%
   Y.~B. Zel'dovich, ~I.~D. Novikov,
   Sov. Astron. volume 10, pages 602, 1967.
%  \bibAnnoteFile{NoStop}{1967SvA....10..602Z}%

%\cite{Carr:1974nx}
%\bibitem{Carr:1974nx} 
  B.~J.~Carr and S.~W.~Hawking,
  %``Black holes in the early Universe,''
  Mon.\ Not.\ Roy.\ Astron.\ Soc.\  {\bf 168}, 399 (1974).
  %%CITATION = MNRAA,168,399;%%
  %397 citations counted in INSPIRE as of 08 Dec 2016

%\cite{Hawking:1974rv}
\bibitem{Hawking:1974rv} 
  S.~W.~Hawking,
  %``Black hole explosions,''
  Nature {\bf 248}, 30 (1974).
  doi:10.1038/248030a0
  %%CITATION = doi:10.1038/248030a0;%%
  %2321 citations counted in INSPIRE as of 08 Dec 2016

%\cite{Georg:2016yxa}
\bibitem{Georg:2016yxa} 
  J.~Georg, G.~Sengor, and S.~Watson,
  %``Nonthermal WIMPs and primordial black holes,''
  Phys.\ Rev.\ D {\bf 93}, no. 12, 123523 (2016)
  doi:10.1103/PhysRevD.93.123523
  [arXiv:1603.00023 [hep-ph]].
  %%CITATION = doi:10.1103/PhysRevD.93.123523;%%
  %2 citations counted in INSPIRE as of 05 Dec 2016
  
 %\cite{Harada:2016mhb}
\bibitem{Harada:2016mhb} 
  T.~Harada, C.~M.~Yoo, K.~Kohri, K.~i.~Nakao and S.~Jhingan,
  %``Primordial black hole formation in the matter-dominated phase of the Universe,''
  Astrophys.\ J.\  {\bf 833}, no. 1, 61 (2016)
  doi:10.3847/1538-4357/833/1/61
  [arXiv:1609.01588 [astro-ph.CO]].
  %%CITATION = doi:10.3847/1538-4357/833/1/61;%%
  %2 citations counted in INSPIRE as of 22 Dec 2016 
  
  
%\cite{Carr:2009jm}
\bibitem{Carr:2009jm} 
  B.~J.~Carr, K.~Kohri, Y.~Sendouda and J.~Yokoyama,
  %``New cosmological constraints on primordial black holes,''
  Phys.\ Rev.\ D {\bf 81}, 104019 (2010)
  doi:10.1103/PhysRevD.81.104019
  [arXiv:0912.5297 [astro-ph.CO]].
  %%CITATION = doi:10.1103/PhysRevD.81.104019;%%
  %255 citations counted in INSPIRE as of 29 Nov 2016

  %\cite{Nemiroff:2001bp}
\bibitem{Nemiroff:2001bp} 
  R.~J.~Nemiroff, G.~F.~Marani, J.~P.~Norris and J.~T.~Bonnell,
  %``Limits on the cosmological abundance of supermassive compact objects from a millilensing search in gamma-ray burst data,''
  Phys.\ Rev.\ Lett.\  {\bf 86}, 580 (2001)
  doi:10.1103/PhysRevLett.86.580
  [astro-ph/0101488].
  %%CITATION = doi:10.1103/PhysRevLett.86.580;%%
  %13 citations counted in INSPIRE as of 11 Dec 2016
  
  
%\cite{Griest:2013esa}
\bibitem{Griest:2013esa} 
  K.~Griest, A.~M.~Cieplak and M.~J.~Lehner,
  %``New Limits on Primordial Black Hole Dark Matter from an Analysis of Kepler Source Microlensing Data,''
  Phys.\ Rev.\ Lett.\  {\bf 111}, no. 18, 181302 (2013).
  doi:10.1103/PhysRevLett.111.181302
  %%CITATION = doi:10.1103/PhysRevLett.111.181302;%%
  %34 citations counted in INSPIRE as of 11 Dec 2016  

 %\cite{Ricotti:2007au}
\bibitem{Ricotti:2007au} 
  M.~Ricotti, J.~P.~Ostriker and K.~J.~Mack,
  %``Effect of Primordial Black Holes on the Cosmic Microwave Background and Cosmological Parameter Estimates,''
  Astrophys.\ J.\  {\bf 680}, 829 (2008)
  doi:10.1086/587831
  [arXiv:0709.0524 [astro-ph]].
  %%CITATION = doi:10.1086/587831;%%
  %96 citations counted in INSPIRE as of 11 Dec 2016 

%\cite{Chen:2016pud}
\bibitem{Chen:2016pud} 
  L.~Chen, Q.~G.~Huang and K.~Wang,
  %``Constraint on the abundance of primordial black holes in dark matter from Planck data,''
  arXiv:1608.02174 [astro-ph.CO].
  %%CITATION = ARXIV:1608.02174;%%
  %7 citations counted in INSPIRE as of 05 Dec 2016
  
%\cite{Ali-Haimoud:2016mbv}
\bibitem{Ali-Haimoud:2016mbv} 
  Y.~Ali-Hamoud and M.~Kamionkowski,
  %``Cosmic microwave background limits on accreting primordial black holes,''
  arXiv:1612.05644 [astro-ph.CO].
  %%CITATION = ARXIV:1612.05644;%%  

%\cite{Gaggero:2016dpq}
\bibitem{Gaggero:2016dpq} 
  D.~Gaggero, G.~Bertone, F.~Calore, R.~M.~T.~Connors, M.~Lovell, S.~Markoff and E.~Storm,
  %``Searching for Primordial Black Holes in the radio and X-ray sky,''
  arXiv:1612.00457 [astro-ph.HE].
  %%CITATION = ARXIV:1612.00457;%%

%\cite{Carr:2016drx}
\bibitem{Carr:2016drx} 
  B.~Carr, F.~Kuhnel and M.~Sandstad,
  %``Primordial Black Holes as Dark Matter,''
  Phys.\ Rev.\ D {\bf 94}, no. 8, 083504 (2016)
  doi:10.1103/PhysRevD.94.083504
  [arXiv:1607.06077 [astro-ph.CO]].
  %%CITATION = doi:10.1103/PhysRevD.94.083504;%%
  %18 citations counted in INSPIRE as of 29 Nov 2016
 
 %\cite{Green:2016xgy}
\bibitem{Green:2016xgy} 
  A.~M.~Green,
  %``Microlensing and dynamical constraints on primordial black hole dark matter with an extended mass function,''
  Phys.\ Rev.\ D {\bf 94}, no. 6, 063530 (2016)
  doi:10.1103/PhysRevD.94.063530
  [arXiv:1609.01143 [astro-ph.CO]].
  %%CITATION = doi:10.1103/PhysRevD.94.063530;%%
  %5 citations counted in INSPIRE as of 11 Dec 2016
 
%\cite{Aghanim:2015xee}
\bibitem{Aghanim:2015xee} 
  N.~Aghanim {\it et al.} [Planck Collaboration],
  %``Planck 2015 results. XI. CMB power spectra, likelihoods, and robustness of parameters,''
  Astron.\ Astrophys.\  {\bf 594}, A11 (2016)
  doi:10.1051/0004-6361/201526926
  [arXiv:1507.02704 [astro-ph.CO]].
  %%CITATION = doi:10.1051/0004-6361/201526926;%%
  %205 citations counted in INSPIRE as of 10 Dec 2016  

%\cite{Poulin:2016anj}
\bibitem{Poulin:2016anj} 
  V.~Poulin, J.~Lesgourgues and P.~D.~Serpico,
  %``Cosmological constraints on exotic injection of electromagnetic energy,''
  arXiv:1610.10051 [astro-ph.CO].
  %%CITATION = ARXIV:1610.10051;%%}.


%\cite{Tashiro:2008sf}
\bibitem{Tashiro:2008sf} 
  H.~Tashiro and N.~Sugiyama,
  %``Constraints on Primordial Black Holes by Distortions of Cosmic Microwave Background,''
  Phys.\ Rev.\ D {\bf 78}, 023004 (2008)
  doi:10.1103/PhysRevD.78.023004
  [arXiv:0801.3172 [astro-ph]].
  %%CITATION = doi:10.1103/PhysRevD.78.023004;%%
  %19 citations counted in INSPIRE as of 11 Dec 2016
  

%\cite{Zhang:2007zzh}
\bibitem{Zhang:2007zzh} 
  L.~Zhang, X.~Chen, M.~Kamionkowski, Z.~g.~Si and Z.~Zheng,
  %``Constraints on radiative dark-matter decay from the cosmic microwave background,''
  Phys.\ Rev.\ D {\bf 76}, 061301 (2007)
  doi:10.1103/PhysRevD.76.061301
  [arXiv:0704.2444 [astro-ph]].
  %%CITATION = doi:10.1103/PhysRevD.76.061301;%%
  %91 citations counted in INSPIRE as of 19 Dec 2016

%\cite{Madhavacheril:2013cna}
\bibitem{Madhavacheril:2013cna} 
  M.~S.~Madhavacheril, N.~Sehgal and T.~R.~Slatyer,
  %``Current Dark Matter Annihilation Constraints from CMB and Low-Redshift Data,''
  Phys.\ Rev.\ D {\bf 89}, 103508 (2014)
  doi:10.1103/PhysRevD.89.103508
  [arXiv:1310.3815 [astro-ph.CO]].
  %%CITATION = doi:10.1103/PhysRevD.89.103508;%%
  %92 citations counted in INSPIRE as of 29 Nov 2016

%\cite{Slatyer:2015jla}
\bibitem{Slatyer:2015jla} 
  T.~R.~Slatyer,
  %``Indirect dark matter signatures in the cosmic dark ages. I. Generalizing the bound on s-wave dark matter annihilation from Planck results,''
  Phys.\ Rev.\ D {\bf 93}, no. 2, 023527 (2016)
  doi:10.1103/PhysRevD.93.023527
  [arXiv:1506.03811 [hep-ph]].
  %%CITATION = doi:10.1103/PhysRevD.93.023527;%%
  %49 citations counted in INSPIRE as of 29 Nov 2016

%\cite{Slatyer:2015kla}
\bibitem{Slatyer:2015kla} 
  T.~R.~Slatyer,
  %``Indirect Dark Matter Signatures in the Cosmic Dark Ages II. Ionization, Heating and Photon Production from Arbitrary Energy Injections,''
  Phys.\ Rev.\ D {\bf 93}, no. 2, 023521 (2016)
  doi:10.1103/PhysRevD.93.023521
  [arXiv:1506.03812 [astro-ph.CO]].
  %%CITATION = doi:10.1103/PhysRevD.93.023521;%%
  %15 citations counted in INSPIRE as of 29 Nov 2016

%\cite{Liu:2016cnk}
\bibitem{Liu:2016cnk} 
  H.~Liu, T.~R.~Slatyer and J.~Zavala,
  %``Contributions to cosmic reionization from dark matter annihilation and decay,''
  Phys.\ Rev.\ D {\bf 94}, no. 6, 063507 (2016)
  doi:10.1103/PhysRevD.94.063507
  [arXiv:1604.02457 [astro-ph.CO]].
  %%CITATION = doi:10.1103/PhysRevD.94.063507;%%
  %5 citations counted in INSPIRE as of 29 Nov 2016

%\cite{Slatyer:2016qyl}
\bibitem{Slatyer:2016qyl} 
  T.~R.~Slatyer and C.~L.~Wu,
  %``General Constraints on Dark Matter Decay from the Cosmic Microwave Background,''
  arXiv:1610.06933 [astro-ph.CO].
  %%CITATION = ARXIV:1610.06933;%%


%\cite{MacGibbon:1991tj}
\bibitem{MacGibbon:1991tj} 
%\cite{MacGibbon:1990zk}
%\bibitem{MacGibbon:1990zk} 
  J.~H.~MacGibbon and B.~R.~Webber,
  %``Quark and gluon jet emission from primordial black holes: The instantaneous spectra,''
  Phys.\ Rev.\ D {\bf 41}, 3052 (1990).
  doi:10.1103/PhysRevD.41.3052;
  %%CITATION = doi:10.1103/PhysRevD.41.3052;%%
  %151 citations counted in INSPIRE as of 29 Nov 2016

  J.~H.~MacGibbon,
  %``Quark and gluon jet emission from primordial black holes. 2. The Lifetime emission,''
  Phys.\ Rev.\ D {\bf 44}, 376 (1991).
  doi:10.1103/PhysRevD.44.376
  %%CITATION = doi:10.1103/PhysRevD.44.376;%%
  %79 citations counted in INSPIRE as of 29 Nov 2016

%\cite{Bolton:2011ck}
\bibitem{Bolton:2011ck} 
  J.~S.~Bolton, G.~D.~Becker, S.~Raskutti, J.~S.~B.~Wyithe, M.~G.~Haehnelt and W.~L.~W.~Sargent,
  %``Improved measurements of the intergalactic medium temperature around quasars: possible evidence for the initial stages of He-II reionisation at z~6,''
  doi:10.1111/j.1365-2966.2011.19929.x
  arXiv:1110.0539 [astro-ph.CO].
  %%CITATION = doi:10.1111/j.1365-2966.2011.19929.x;%%
  %3 citations counted in INSPIRE as of 22 Dec 2016

%\cite{Bolton:2010gr}
\bibitem{Bolton:2010gr} 
  J.~S.~Bolton, G.~D.~Becker, J.~S.~B.~Wyithe, M.~G.~Haehnelt and W.~L.~W.~Sargent,
  %``A first direct measurement of the intergalactic medium temperature around a quasar at z=6,''
  Mon.\ Not.\ Roy.\ Astron.\ Soc.\  {\bf 406}, 612 (2010)
  doi:10.1111/j.1365-2966.2010.16701.x
  [arXiv:1001.3415 [astro-ph.CO]].
  %%CITATION = doi:10.1111/j.1365-2966.2010.16701.x;%%
  %29 citations counted in INSPIRE as of 22 Dec 2016


%\cite{Belotsky:2014twa}
\bibitem{Belotsky:2014twa} 
  K.~M.~Belotsky and A.~A.~Kirillov,
  %``Primordial black holes with mass $10^{16}-10^{17}$ g and reionization of the Universe,''
  JCAP {\bf 1501}, no. 01, 041 (2015)
  doi:10.1088/1475-7516/2015/01/041
  [arXiv:1409.8601 [astro-ph.CO]].
  %%CITATION = doi:10.1088/1475-7516/2015/01/041;%%
  %5 citations counted in INSPIRE as of 19 Dec 2016

%\cite{Slatyer:2012yq}
\bibitem{Slatyer:2012yq} 
  T.~R.~Slatyer,
  %``Energy Injection And Absorption In The Cosmic Dark Ages,''
  Phys.\ Rev.\ D {\bf 87}, no. 12, 123513 (2013)
  doi:10.1103/PhysRevD.87.123513
  [arXiv:1211.0283 [astro-ph.CO]].
  %%CITATION = doi:10.1103/PhysRevD.87.123513;%%
  %53 citations counted in INSPIRE as of 29 Nov 2016

%\cite{AliHaimoud:2010dx}
\bibitem{AliHaimoud:2010dx} 
  Y.~Ali-Haimoud and C.~M.~Hirata,
  %``HyRec: A fast and highly accurate primordial hydrogen and helium recombination code,''
  Phys.\ Rev.\ D {\bf 83}, 043513 (2011)
  doi:10.1103/PhysRevD.83.043513
  [arXiv:1011.3758 [astro-ph.CO]].
  %%CITATION = doi:10.1103/PhysRevD.83.043513;%%
  %52 citations counted in INSPIRE as of 07 Dec 2016

%\cite{Lewis:1999bs}
\bibitem{Lewis:1999bs} 
  A.~Lewis, A.~Challinor and A.~Lasenby,
  %``Efficient computation of CMB anisotropies in closed FRW models,''
  Astrophys.\ J.\  {\bf 538}, 473 (2000)
  doi:10.1086/309179
  [astro-ph/9911177].
  %%CITATION = doi:10.1086/309179;%%
  %2104 citations counted in INSPIRE as of 10 Dec 2016
  
%\cite{Howlett:2012mh}
\bibitem{Howlett:2012mh} 
  C.~Howlett, A.~Lewis, A.~Hall and A.~Challinor,
  %``CMB power spectrum parameter degeneracies in the era of precision cosmology,''
  JCAP {\bf 1204}, 027 (2012)
  doi:10.1088/1475-7516/2012/04/027
  [arXiv:1201.3654 [astro-ph.CO]].
  %%CITATION = doi:10.1088/1475-7516/2012/04/027;%%
  %66 citations counted in INSPIRE as of 10 Dec 2016

%\cite{Lewis:2013hha}
\bibitem{Lewis:2013hha} 
  A.~Lewis,
  %``Efficient sampling of fast and slow cosmological parameters,''
  Phys.\ Rev.\ D {\bf 87}, no. 10, 103529 (2013)
  doi:10.1103/PhysRevD.87.103529
  [arXiv:1304.4473 [astro-ph.CO]].
  %%CITATION = doi:10.1103/PhysRevD.87.103529;%%
  %81 citations counted in INSPIRE as of 10 Dec 2016

%\cite{Lewis:2002ah}
\bibitem{Lewis:2002ah} 
  A.~Lewis and S.~Bridle,
  %``Cosmological parameters from CMB and other data: A Monte Carlo approach,''
  Phys.\ Rev.\ D {\bf 66}, 103511 (2002)
  doi:10.1103/PhysRevD.66.103511
  [astro-ph/0205436].
  %%CITATION = doi:10.1103/PhysRevD.66.103511;%%
  %1883 citations counted in INSPIRE as of 10 Dec 2016

%\cite{Padmanabhan:2005es}
\bibitem{Padmanabhan:2005es} 
  N.~Padmanabhan and D.~P.~Finkbeiner,
  %``Detecting dark matter annihilation with CMB polarization: Signatures and experimental prospects,''
  Phys.\ Rev.\ D {\bf 72}, 023508 (2005)
  doi:10.1103/PhysRevD.72.023508
  [astro-ph/0503486].
  %%CITATION = doi:10.1103/PhysRevD.72.023508;%%
  %163 citations counted in INSPIRE as of 18 Dec 2016


%\cite{PlanckTable:2015jan}
\bibitem{PlanckTable:2015jan}
  ``Planck 2015 Results: Cosmological Parameter Tables,''
  \nolinkurl{https://wiki.cosmos.esa.int/planckpla2015/images/6/67/Params_table_2015_limit68.pdf}
    

%\cite{Weidenspointner:1999}
\bibitem{Weidenspointner:1999}   
  G. Weidenspointner. 1999. The Origin of the Cosmic Gamma-Ray Background in the COMPTEL Energy Range. PhD thesis, Technical University of Munich, Munich, Germany ?id=602832. \nolinkurl{http://mediatum.ub.tum.de}

%\cite{Zavala:2009mi}
\bibitem{Zavala:2009mi} 
  J.~Zavala, M.~Vogelsberger and S.~D.~M.~White,
  %``Relic density and CMB constraints on dark matter annihilation with Sommerfeld enhancement,''
  Phys.\ Rev.\ D {\bf 81}, 083502 (2010)
  doi:10.1103/PhysRevD.81.083502
  [arXiv:0910.5221 [astro-ph.CO]].
  %%CITATION = doi:10.1103/PhysRevD.81.083502;%%
  %66 citations counted in INSPIRE as of 29 Nov 2016

 %\cite{Strigari:2005hu}
\bibitem{Strigari:2005hu} 
  L.~E.~Strigari, J.~F.~Beacom, T.~P.~Walker and P.~Zhang,
  %``The Concordance Cosmic Star Formation Rate: Implications from and for the supernova neutrino and gamma ray backgrounds,''
  JCAP {\bf 0504}, 017 (2005)
  doi:10.1088/1475-7516/2005/04/017
  [astro-ph/0502150].
  %%CITATION = doi:10.1088/1475-7516/2005/04/017;%%
  %79 citations counted in INSPIRE as of 08 May 2016

%\cite{Ruiz-Lapuente:2015yua}
\bibitem{Ruiz-Lapuente:2015yua} 
  P.~Ruiz-Lapuente, L.~S.~The, D.~Hartmann, M.~Ajello, R.~Canal, F.~K.~Ropke, S.~T.~Ohlmann and W.~Hillebrandt,
  %``The origin of the cosmic gamma-ray background in the MeV range,''
  Astrophys.\ J.\  {\bf 820}, no. 2, 142 (2016)
  doi:10.3847/0004-637X/820/2/142
  [arXiv:1502.06116 [astro-ph.HE]].
  %%CITATION = doi:10.3847/0004-637X/820/2/142;%%
  %2 citations counted in INSPIRE as of 08 May 2016

%\cite{Horiuchi:2010kq}
\bibitem{Horiuchi:2010kq} 
  S.~Horiuchi and J.~F.~Beacom,
  %``Revealing Type Ia supernova physics with cosmic rates and nuclear gamma rays,''
  Astrophys.\ J.\  {\bf 723}, 329 (2010)
  doi:10.1088/0004-637X/723/1/329
  [arXiv:1006.5751 [astro-ph.CO]].
  %%CITATION = doi:10.1088/0004-637X/723/1/329;%%
  %38 citations counted in INSPIRE as of 21 Dec 2016

%\cite{DeAngelis:2016slk}
\bibitem{DeAngelis:2016slk} 
  A.~De Angelis {\it et al.} [e-ASTROGAM Collaboration],
  %``The e-ASTROGAM mission (exploring the extreme Universe with gamma rays in the MeV-GeV range),''
  arXiv:1611.02232 [astro-ph.HE].
  %%CITATION = ARXIV:1611.02232;%%
  %1 citations counted in INSPIRE as of 21 Dec 2016

\end{thebibliography}
\end{document}